\documentclass{llncs}
\pdfoutput=1
\usepackage{graphicx}
\usepackage{amsfonts, amsmath,amscd}
\usepackage{mathtools}
\usepackage{amssymb}
\usepackage{tikz-cd}
\usepackage{algorithm}
\usepackage{algorithmic}
\DeclareRobustCommand{\stirling}{\genfrac\{\}{0pt}{}}
\newtheorem{observation}{Observation}
\newtheorem{statement}{Statement}

\title{Semi-Countable Sets and their Application to Search Problems}
\author{Pieter Adriaans\inst{1}}

\institute{ILLC, FNWI-IVI, SNE\\
University of Amsterdam,\\Science Park 107\\ 1098 XG Amsterdam, \\The
Netherlands.
\email{P.W.Adriaans@uva.nl}}

\date{}
\begin{document}
\maketitle

\begin{abstract}
We present the concept of the \emph{information efficiency of functions} as a technique to understand the interaction between information and computation. Based on these results we identify a new class of objects that we call \emph{Semi-Countable Sets}. As the name suggests these sets form a separate class of objects between countable and uncountable sets. In principle these objects are countable, but the information in the descriptions of the elements of the class grows faster than the information in the natural numbers that index them. Any characterization of the class in terms of natural numbers is fundamentally incomplete. Semi-countable sets define one-to-one  injections into the set of natural numbers that can be computed in exponential time, but not in polynomial time. A characteristic semi-countable object is $\phi_{\Sigma}$ the set of all additions for all finite sets of natural numbers.  The class $\phi_{\Sigma}$ codes the Subset Sum problem. This gives a natural and transparant analysis of the separation between the classes $P$ and $NP$.

\end{abstract}

\section{Introduction}

This paper develops some ideas that were presented in an elementary form in  \cite{adri2018-1}, where we argued that the most urgent problem of modern philosophy of information was our lack of understanding of the \emph{interaction between information and computation}. For a deeper understanding of the philsophical backgrounds we refer to this publication.

\subsection{Informal presentation of the main argument \label{MA}}

Interaction between information and computation is  a phenomenon that we are all familiar with from a cognitive point of view, but that until now has eluded mathematical conceptualization. Suppose we want to add a set of numbers $s_1= \{3,4,6,7 \}$. We could compute $((3 + 4) + 6) + 7=20$, but most of us would immediately see that is easier to compute $(3 + 7)+(4+6)= 10 + 10 = 20$. Such a trick is not available for the set  $s_2 =  \{2,5,6,7 \}$. This example shows that, for some sets, the sequence of our computations influences the hardness of the problem. The observation that $\Sigma_{i \in s_1}i = 20$ is, from this perspective,  \emph{less surprising} than the fact that  $\Sigma_{i \in s_2}i = 20$. From an information theoretical point of view this implies that the statement $\Sigma_{i \in s_1}i = 20$ contains \emph{less information} than the statement $\Sigma_{i \in s_2}i = 20$. Since addition is commutative and associative we have no mathematical tools to explain this phenomenon in classical arithmetic.~\footnote{For an elaborate analysis of this example, consult the last part of the Appendix in paragraph \ref{A1}.} 

What we need is a theory that helps us to distinguish the information aspects of different computational histories. Below we develop such a theory. The central concept is the notion of the \emph{information efficiency of a function} as the balance between the information in the input and the information in the output. For addition this gives $\delta(x + y) = \log(x + y) - \log x - \log y$. It is clear that this operation is not associative $\delta((a +b) + c) \neq  \delta(a +(b + c))$. The amount of possible computational histories for addition of a set of numbers is super exponential in its cardinality. This implies that the computational history of the way the output is computed is relevant for the amount of information it contains conditional to the input. In other words: even if we have the answer, we don't know what we know, until we know how it is computed.

For some types of problems this means that, knowing the answer does not help us much to reconstruct the problem. A typical example is the so-called Subset Sum Problem: given a set of natural numbers $s$, is there a subset $s_i \subseteq s$ that adds up to $k$?  Now consider the following statement:

\begin{statement}
 $a$ is the $n$-th  subset of $S$ that adds up to $k$.   
\end{statement}

Here $''a''$ is the name of a set and ``the $n$-th subset of $S$ that adds up to $k$'' a unique description. Note that we can compute the unique description effectively when we have the name and vice versa: i.e. there exists a computable bijection between the set of \emph{names} and the set of \emph{unique descriptions}.  We have an algorithm to solve the search problem corresponding with the statement and the associated decision problem effectively: 

\begin{enumerate}
\item Search problem: What is  the $n$-th subset of $S$ that adds up to $k$? 
\item Decision problem: Does the $n$-th subset of $S$ that adds up to $k$ exist?
\end{enumerate}

A central question is: 

\begin{question}\label{Q1}
Are there uniquely identifying descriptions of objects that contain more or less information than the names of the objects they denote?
\end{question}

The prima facie answer to this question is no. If we can compute the name from the description and vice versa, independent of the amount of time this takes, the descriptive complexities should stay close to each other in the limit. 

On the other hand, observe that computable bijections are by definition information efficient. When the information efficiency of a function is not well-defined, the bijection is also not well defined. We call this the \emph{Principle of Characteristic Information Efficiency}: if two computations have a different information efficiency, then different functions are involved in their computation. As we have seen this is the case for addition.  In this particular case there is no single finite mathematical function that describes the information efficiency of the bijection between sets and the sums of their subsets. There are infinitely many in the limit. The unique description ``the first subset of $s$ that adds up to $k$'' is \emph{ad hoc}. It has no clear relation with the name of the denoted subset $a$. This kind of ad hoc unique descriptions are abundant in every day life. Especially in relation to expressions like ``the first $x$ that ... '' or ``the $n$-th $x$ that ...''. Consider the descriptions: 

\begin{enumerate}
\item The first four-leaf clover I'll find this afternoon. 
\item The $25$-th man with a moustache I'll see in the city. 
\end{enumerate}

It is clear the description ``The first four-leaf clover I'll find this afternoon'' does not describe an intrinsic property of a certain plant. Actually, which plant I'll find (if I find one) completely depends on my search method. The descriptive complexity, of the plant from my perspective at the moment I utter the phrase is a combination of the amount of information in the description itself and the description of the search method I'm going to use. Suppose I throw a dice  to select my path when roaming around the city to find guys with moustaches,  then the description the search process to identify  ``The $25$-th man with a moustache I'll see in the city'' can easily contain more information than the descriptive complexity of his name. 

According to this analysis the answer to research question \ref{Q1} is positive: if a description of the search process, plus a partial description of the object, determines the object we will find, then this description possibly is more complex than the name of the object itself. At the same time the unique description itself can contain much less information.  Mutatis mutandis, there is no way that we can search systematically for the ``the first subset of $s$ that adds up to $k$'' using the information given in $k$. A fortiori there is no search process that works in time polynomial to the complexity of the search problem. The search process creates the object I will find. 

  Ofcourse, given the number $k$ we can always find a subset that adds up to $k$ by simply enumerating all possible subsets ordered by cardinality and compute the sum. In this case we are \emph{not using} the specific information given in $k$ in the \emph{organization} of the search process. Such a \emph{search by enumeration} simply generates the missing information about $s_1$ given $s$ by a process of counting.  
  
A question that emerges is whether there are sets for which there is no faster way to find a solution than by pure enumeration of the possible solutions. We prove this for a (prima facie) relatively simple countable object: the set of natural numbers $\mathbb{N}$ and its finite subsets. We write this set as $\mathfrak{P}(\mathbb{N})$.  We investigate two mappings to a two dimensional infinite discrete space $\mathbb{N}^2$ (think of a chess board that extends to infinity on  two sides): 

 \begin{itemize}
\item We show (via an elaborate counting argument) that  $\mathfrak{P}(\mathbb{N})$  can be mapped efficiently onto $\mathbb{N}^2$. In every cell there is exactly one finite set of natural numbers and vice versa. 
 \item We observe that all possible descriptions of the form ``the $n$-th subset of $\mathbb{N}$ that adds up to $k$'' can be mapped trivially onto the plane $\mathbb{N}^2$: here $n$ is a column and $k$ is a row. In every cell there is exactly one unique description of a set.  Not all unique descriptions will denote, e.g. ``the $10$-th set that adds up to $2$'' does not exist. The infinite space $\mathbb{N}^2$ is compressed infinitely over the $y$-axis. 
  \end{itemize} 
  
Comparable mappings form the core of Cantor's argument that proves the existence of superfinite sets. Our argument follows a related strategy. The Cantor packing function maps the set of natural numbers $\mathbb{N}$ onto the discrete plane $\mathbb{N}^2$.  Using this construction we can investigate all possible mappings between sets and their descriptions in terms of elastic translations over the two dimensional space. We show that all possible mappings between names of sets and their descriptions are \emph{unboundedly information expanding}. The description of most typical sets $s$ as ``the $n$-th subset of $\mathbb{N}$ that adds up to $k$'' using the numbers $n$ and $k$ contains more information than the index of $s$ itself. Another way of formulating this insight is that the information in the natural numbers does  not grow fast enough to characterize the set of descriptions. The set of descriptions is \emph{semi-countable}: their complexity ``outruns'' the information in the set of natural numbers unboundedly in the limit. The set $\mathbb{N}$ is not rich enough to describe semi-countable sets. Consequently search by enumeration is the fastest way to construct them algorithmically.

\subsection{Overview of the paper}
We start with a conceptual overview of various types of computational processes: primitive recursive, $\mu$-recursive and non-deterministic. We show that $\mu$-recursive processes have a special status in so far that they allow for unbounded counting. We show that these processes generate information in logarithmic time in special circumstances. 

We analyze this insight in the context of Kolmogorov complexity and Levin complexity and observe that these measures are not accurate enough for our purpose. We propose the concept of \emph{Information Efficiency of functions} als an alternative complexity measurement theory. We give a detailed analysis of the recursive functions. We observe that  information efficiency is not associative for addition. 

 We study the Cantor pairing function as an information preserving bijection between $\mathbb{N}$ and $\mathbb{N}^2$.  We study the information efficiency of elastic  translations over Cantor bijections over the space $\mathbb{N}^2$. We show that there is a spectrum of these translations. 
 
 We show that the set $\mathfrak{P}(\mathbb{N})$, of all finite sets of natural numbers,  can be mapped onto  $\mathbb{N}^2$ efficiently. This allows us to investigate the general conditions for elastic translations based on addition and multiplication of sets of numbers. The object that describes all possible additions of finite sets of natural numbers is $\phi_{\Sigma}$. The corresponding object for multiplication is $\phi_{\Pi}$.  We show that the resulting set of unique descriptions of sets ``the $n$-th subset of $\mathbb{N}$ that adds up to $k$'' is 
 \begin{enumerate}
 \item associated with an infinite number of different computations. 
 \item not fully characterized by the set of natural numbers: the information in the descriptions grows faster than any counting process.  
 \end{enumerate}
 The set $\phi_{\Sigma}$ is semi-countable. We can search sets in exponential time but not in polynomial time. The object $\phi_{\Pi}$ is fundamentally less complex as a result of the fact that the information efficiency for multiplication is associative. 

This argument can easily be generalized to the Subset Sum problem, which proves the separation between $N$ and $NP$.

\section{Conceptual Analysis}

In this paragraph we give a conceptual analysis of the issues concerning the interaction between information and computation. For a more global discussion of the underlying philosophical problems the reader is referred to \cite{adri2018-1}. We will use the prefix free Kolmogorov complexity $K(x)$ as our measure of descriptive complexity of a string $x$ and \cite{LiVi08} as basic reference: $K(x)$ is the length of the shortest program that computes $x$ on a reference Universal Turing machine. 

\subsection{Types of Computational Processes}
 
There are at least three fundamentally different types of computing (See Figure \ref{Computing-Classes}) :

\begin{figure}[ht!]
\centering
\includegraphics[width=70mm]{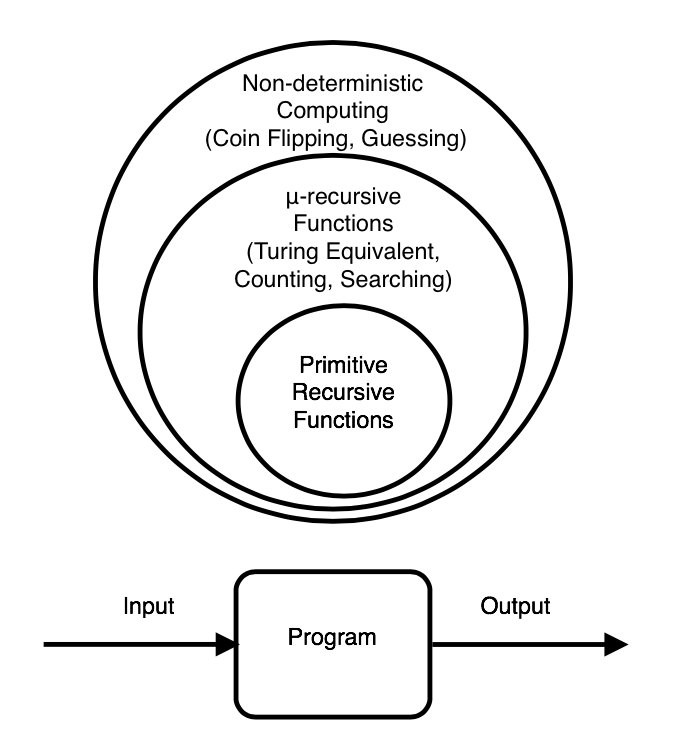}
\caption{Classes of Computing Systems \label{Computing-Classes}}
\end{figure}

\begin{itemize}
\item \emph{Elementary deterministic computing} as embodied in the primitive recursive functions. This kind of computing  does not generate information: the amount of information in the Output is limited by the sum of the descriptive complexity of the Input and the Program.  
\item \emph{Deterministic computing enriched with search} (bounded or unbounded) as embodied by the class of Turing equivalent systems, specifically the $\mu$-recursive functions. This type of computing  generates information at logarithmic speed: the amount of information in the Output is not limited by the sum of the descriptive complexities of the Input and the Program. 
\item \emph{Non-deterministic computing generates} information at linear speed. 
\end{itemize}

Suppose there is a class of search problems with a polynomial time checking function that cannot be solved by a deterministic program but  can be solved using bounded search. Such a search routine would take exponential time, since information generation has logarithmic speed. A non-deterministic computer could generate (guess) the required information at linear speed and then perform the test in polynomial time. The existence of such a class of search problems would indicate a separation between $P$ and $NP$ for the associated decision problems: these problems cannot be solved deterministically, they can be solved using bounded search in exponential time and non-deterministically in polynomial time. 
 
This analysis shows that the formulation of search problems in terms of polynomial time bounds and Turing Machines might be quite misleading. More important than the polynomial time bound is the fact that the search functions can not be computed \emph{at all} (in general) by deterministic functions, while the distinction between deterministic search and primitive recursion is hard to make in the context of Turing machines. We analyse this issue in the following paragraph.

\subsection{The $\mu$-operator for unbounded search}

There is a subtle difference between systematic search and deterministic construction that is blurred in our current definitions of what computing is. If one considers the three fundamental equivalent theories of computation, Turing machines, $\lambda$-calculus and recursion theory, only the latter defines a clear distinction between construction and search, in terms of the difference between primitive recursive functions and $\mu$-recursive functions. The  set of primitive recursive functions consists of: the  constant function, the successor function, the projection function, composition and primitive recursion. With these we can define everyday mathematical functions like addition, subtraction, multiplication, division, exponentiation etc. In order to get full Turing equivalence one must add the $\mu$-operator. In the world of Turing machines this device coincides with infinite loops associated with undefined variables. It is defined as follows in \cite{Odi16}:

 For every 2-place function $f(x,y)$ one can define a new function, $g(x) = \mu y[f(x,y)=0 ] $, where $g(x)$ returns the smallest number y such that $f(x,y) = 0.$  Defined in this way $\mu$ is a partial function. One way to think about $\mu$ as in terms of an operator that tries to compute in succession all the values $f(x,0)$, $f(x,1)$, $f(x,2)$, ... until for some $m$ $f(x,m)$ returns $0$, in which case such an $m$ is returned. In this interpretation, if $m$ is the first value for which $f(x,m) = 0 $ and thus $g(x) = m$, the expression  $\mu y[f(x,y)=0 ]$ is associated with a routine that performs exactly  $m$ successive test computations of the form $f(x,y)$ before finding $m$. Since the $\mu$-operator is unbounded $m$ can have any value. 

Note that the name $g$ does not refer to a function but to a function-scheme. The $x$ in the expression $g(x)$ is not an argument of a function but the index of a function name $f_x(y) \Leftrightarrow f(x,y)$. We can interpret the $\mu$-operator as a meta-operator that has access to an infinite number of primitive recursive functions. In this interpretation there is no such thing as a general search routine.  Each search function is specific: searching for your glasses is different from searching for your wallet, even when you look for them in the same places. 

The difference between primitive recursion and $\mu$-recursion formally defines the difference between \emph{construction} and \emph{search}. Systematic search involves an \emph{enumeration} of all the elements in the search space together with \emph{checking function} that helps us to decide that we have found what we are looking for.  We will have to look into this conception of enumeration in more depth. 

\begin{figure}[ht!]
\centering
\includegraphics[width=120mm]{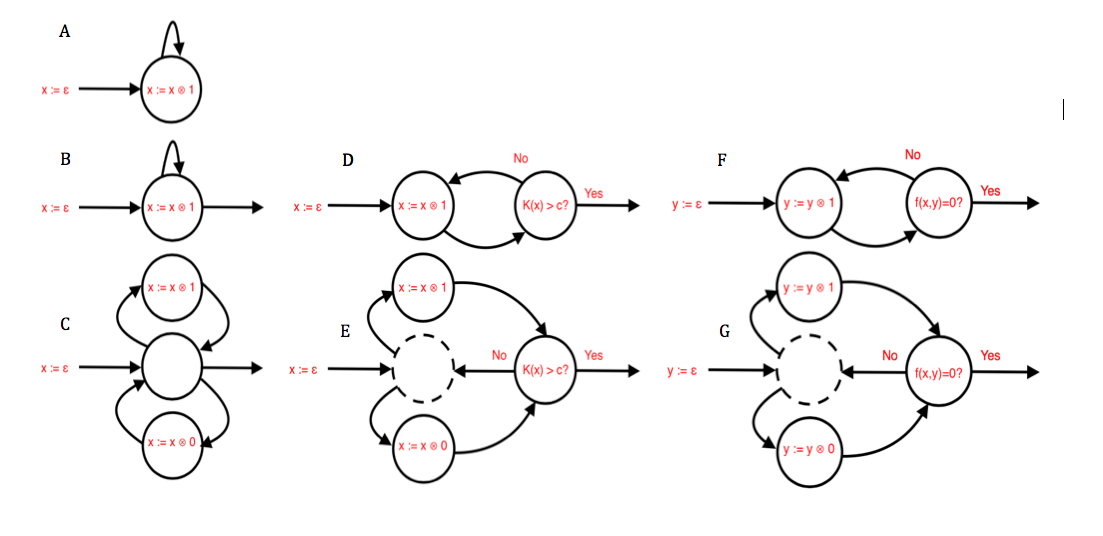}
\caption{Counting Processes \label{Counting-Processes}}
\end{figure}

\subsection{Determinism versus non-determinism \label{NonDetvsDet}}
In this paragraph we discuss the hybrid nature of unary counting processes, which, in a manner of speaking, are positioned between fully deterministic and non-deterministic processes. By definition deterministic processes do not generate new information, because the outcome of the process is determined. For a full discussion of this issue see \cite{adri2018-1}.  We start with a detailed analysis of the seven elementary counting processes (A-G) shown in figure \ref{Counting-Processes}. The tensor operation $\otimes$ signifies concatenation. 
\begin{itemize}
\item Automaton $A$ is deterministic and it does not halt. It starts with an empty string $\epsilon$ and writes an infinite sequence of ones.  
\item Automaton $B$ is non-deterministic. It generates the set of all finite strings of ones, i.e. the set of all finite unary numbers. We will call this a \emph{Counting Automaton} or CA. 
\item Automaton $C$, also known as the \emph{Coin Flipping Automaton} or CFA,  is non-deterministic. It generates the set of all finite binary strings consisting of zeros and ones, i.e. the set of all finite binary numbers.
\item Automaton $D$ is deterministic. It is equivalent to automaton $B$ with the addition of an extra test that checks the Kolmogorov complexity of the string $x$ generated so far. As soon as $x$ has a complexity greater than a constant $c$ the process stops and produces output $x$. We will ignore for the sake of argument that the Kolmogorov complexity is not computable and assume that there is some oracle that gives us a decision on the matter. 
\item Automaton $E$ is non-deterministic. It is equivalent to automaton $C$ with the addition of an extra test that checks the Kolmogorov complexity of the string $x$ generated so far. As soon as $x$ has a complexity greater than a constant $c$ the process stops and produces output $x$. 
\item Automaton $F$ is deterministic. It is equivalent to automaton $D$, but now the test routine involves a computable function $f(x,y)=0$ running on an input index $y$. In fact it is an implementation of the central routine of the $\mu$-recursive search process that we discussed in the previous paragraph. It defines $\mu$-recursion based on a Counting Automaton. 
\item Automaton $G$ is non-deterministic. It is equivalent to automaton $E$. Here also the test routine involves a computable function $f(x,y)=0$ running on an input index $y$. It defines $\mu$-recursion based on a Coin Flipping Automaton. 
\end{itemize}

The difference between automaton $A$ and $B$ illustrates the fact that counting is essentially a non-deterministic operation. Automaton $A$ does not effectively generate an object, whereas $B$ generates all finite unary strings. Consequently the amount of information that $B$ generates is unbounded. The information is generated by a sequence of free binary decisions to continue counting followed by one decision to stop the process.  

The decisions to stop and start the process can be seen as meta-decisions that as such are not an intrinsic part of the process. This is illustrated by the fact that as soon as we add a stop criterion in automaton $D$ and $E$ the unary counting process becomes deterministic and the binary string generation keeps its non-deterministic nature. 

The fact that the systems $B$ and $C$ both generate information is illustrated by systems $D$ and $E$. Both stop at the moment when a certain amount of information of size $c$ is generated. Since unary strings code information very inefficiently, and thus have a low Kolmogorov complexity, process $D$ needs to perform at least $2^c$ write operations before it stops, where process $E$ can reach this goal in principle in $c$ steps. In this case the computation  $D$ has exponential time whereas $E$ can work in linear time. The observations we can make on the basis of this analysis are:

\begin{observation}\label{O1}

 \begin{itemize} 
\item Only non-deterministic processes generate information. Deterministic processes by definition do not generate information \cite{AB2011}.
\item Both unary counting and coin flipping are non-deterministic processes. 
\item Counting generates information at logarithmic speed. The time needed to generate a certain amount of  information by means of unary counting, is exponential in the amount of  time a coin flipping automaton needs to generate the same amount of information. 
\end{itemize}
\end{observation}

Unary counting with a stop criterion is hybrid in the sense that it has characteristics of both deterministic and non-deterministic processes. This explains the special status of these kind of processes in recursion theory. Unary counting with a stop criterion is a form of computing that is essentially stronger than standard deterministic computing.  There are many conceptual problems around this notion of computing, one of which is the fact that the descriptive complexity of the computational process at any time during the computation may be much higher than the complexity of the actual output. See \cite{AB2011} for a discussion. The crucial limiting factor is the descriptive complexity of the halting test: $K(x) < c$ for processes $D$ and $E$ and $f(x,y) = 0$ for processes $F$ and $G$.  A central observation in this context is the so-called \emph{non-monotonicity of set theoretical operations}  (see  \cite{adri2018-1}, par. 6.2).


Since $\mu$-recursion is stronger than primitive recursion there will be  classes of search problems that can be solved by $\mu$-recursive functions and not in general by primitive recursive functions. The search process in $\mu$-recursion is driven by counting.  Consequently a non-deterministic version of  $\mu$-recursion using a Coin Flipping Automaton could solve a search problem, i.e. compute the value $m$ for the test function $g(m)$, in time \emph{linear} in the length of representation of the number $m$, while the search process in the classical deterministic $\mu$-recursion, using a Counting Automaton, would take time \emph{exponential} in the length of the representation of $m$, i.e. the value of $m$. Moreover if the time complexity of the computation $f(x,y)$ is polynomial in the length of the input then the fact that a solution can be generated non-deterministically in linear time would be overshadowed by the time complexity of the checking function.


\section{Information Efficiency of Recursive Functions}
Let $x,y,z \in {\mathbb N}$, where ${\mathbb N}$ denotes the natural
numbers and we identify ${\mathbb N}$ and $\{0,1\}^*$ according to
the correspondence
\[(0, \varepsilon ), (1,0), (2,1), (3,00), (4,01), \ldots \]
Here $\varepsilon$ denotes the {\em empty word}. The {\em length}
$\mathit{l}(x)$ of $x$ is the number of bits in the binary string $x$. in the following we will use the logarithm with base $2$ as our standard reference  $\log x = \log_2 x$. 
The standard reference \cite{LiVi08} for the definitions concerning Kolmogorov complexity is followed.
$K$ is the prefix-free Kolmogorov complexity of a binary string. It is defined as:
\begin{definition}\label{KOL}
\[K(x|y)=  \min_i \{\mathit{l}(\overline{i}):U(\overline{i}y)=x\}\]
\end{definition}
i.e. the shortest self-delimiting index of a Turing machine $T_i$ that produces $x$ on input y, where $i \in \{1,2,...\}$ and $y \in \{0,1\}^{\ast}$. Here $\mathit{l}(|\overline{i}|)$ is the length of a self-delimiting code of an index and $U$ is a universal Turing machine that runs program $y$ after interpreting $\overline{i}$. The length of $\mathit{l}(\overline{i})$ is limited for practical purposes by $n +  2\log n + 1$, where $n=|i|$. The actual Kolmogorov complexity of a string is defined as the one-part code that is the result of, what one could call, the:
\begin{definition}[Forcing operation] \label{forcing-operation}
$K(x)= K(x|\varepsilon)$
\end{definition}

According to the classical view the descriptive complexity of the Output of a deterministic computational proces is bounded by the sum of the complexity of the Input and the Program (See Figure \ref{Computing-Classes}):

\begin{equation}\label{E1}
K(Output) \leq K(Input) + K(Program) + O(1)
\end{equation}

Based on the discussion above we formulate the conjecture: 

\begin{conjecture}\label{C2}
Deterministic computational processes generate information at logarithmic speed. 
\end{conjecture}

Acceptance of this conjecture would imply a shift from Kolmogorov complexity ro Levin complexity which takes the influence of the computing time into account:  

\begin{definition}[Levin complexity: Time = Information]
 The Levin complexity of a string $x$ is the sum of the length $\mathit{l}(p)$ and the logarithm of the computation time of the smallest program $p$ that produces $x$ when it runs on a universal Turing machine $U$, noted as $U(p) = x$:
\[K_t(x) = \min_p
\{\mathit{l}(p) + \log(time(p)), U(p) = x\}\]
\end{definition}

The inequality \ref{E1} then becomes: 

\begin{equation}\label{E2}
K_t(Output) \leq K(Input) + K(Program) +\log(time(Program)) + O(1)
\end{equation}

The problem with such a proposal is that our classical proof techniques and information measures are not sensitive enough to observe  the difference between the two measures in practical situations. Information production at logarithmic speed is extremely slow and we will in every day life never sense the way it influences our measurements. The situation is not unlike the one in the theory of relativity. Our measurement of time is affected by our relative speed, but in every day life the speeds at which we travel in relation to the accuracy of our measurement techniques are such that we do not observe these fluctuations. The same holds for the difference between Kolmogorov and Levin complexity. First of all both measures are uncomputable, so we can never present a convincing example illustrating the difference. Secondly, if we estimate the number of computational steps the universe has made since the Big Bang as $10^{123}$, then a deterministic system would have produced about $400$ bits of information in this time span. Even if we could compute the value of $K(Output)$ and $K_t(Output)$  then the asymptotic nature of the measure, reflected in the $O(1)$ parameter, which is related to the descriptive complexity of the reference Universal machines we are considering, does not allow us to reach the accuracy necessary to observe the difference on the time scale of our universe as a whole. So the classical theory of Kolmogorov complexity is of little use to us. We need to develop more advanced information measurement techniques. 

One possibility is to shift our attention from Turing machines to recursive functions. The big advantage of recursive functions is that we have a reliable definition of primitive recursive functions, which relieves us of the burden to select something like a reference Universal machine, which eliminates the asymptotic nature of the measurement theory we get. Secondly we can model the flow of information through computational processes more effectively. We define: 

 \begin{definition}[Information in Natural numbers]\label{FIRSTLAW}
\[\forall (x \in \mathbb{N}) I(x)=\log x\]
\end{definition}

The rationale behind the choice of the log function as an information measure is discussed extensively in  \cite{adri2018-1}. The big advantage of the definition of an information measure using recursive functions is the fact that we can get rid of the asymptotic $O(1)$ factor, since we do not have to relativize over the class of universal machines. We get a theory about compressible numbers, much in line with Kolmogorov complexity, if we introduce the notion of the information efficiency of a function.  The \emph{Information Efficiency} of a function is the difference between the amount of information in the input of a function and the amount of information in the output. We use the shorthand $f(\overline{x})$ for  $f(x_1,x_2,\dots,x_k)$.  We consider functions on the natural numbers. If we measure the amount of information in a number $n$ as: \[I(n) = \log n\] then we can measure the information effect of applying function $f$ to $n$ as: \[I(f(n)) = \log f(n)\] 

This allows us to estimate the \emph{information efficiency} as: \[\delta(f(n)) = I(f(n)) - I(n))\]
More formally: 

\begin{definition}[Information Efficiency of a Function]\label{EFFFUNCTION}
Let $f: \mathbb{N}^k \rightarrow \mathbb{N}$  be a function of $k$ variables.  We have:
\begin{itemize} 
\item the \emph{input information} $I(\overline{x})$ and 
\item the \emph{output information}   $I(f(\overline{x}))$. 
\item The information efficiency of the expression $ f(\overline{x})$ is  
\[\delta(f(\overline{x}))= I(f(\overline{x})) - I(\overline{x})\]
\item A  function $f$ is \emph{information conserving} if $\delta(f(\overline{x}))=0$ i.e. it contains exactly the amount of information in its input parameters, 
\item it is \emph{information discarding} if  $\delta(f(\overline{x}))<0$ and 
\item it has \emph{constant information } if  $\delta(f(\overline{x})) = 0$. 
\item it is \emph{information expanding} if  $\delta(f(\overline{x}))>0$. 
\end{itemize}
\end{definition}

The big advantage of this definition over Kolmogorov complexity is that we can compute the flow of information through functions exactly. In the Appendix I in paragraph  \ref{A1} we give extensive examples of the information efficiency of elementary recursive functions.

 \begin{definition}[Principle of Characteristic Information Efficiency]\label{InfEff}
 The concept of information efficiency is \emph{characteristic} for a function. Consequently if the information efficiency varies over sets of computable numbers, then different functions must be involved in their computation. 
 \end{definition}
  
    The concept of Information Efficiency gives us a tool to decide between Kolmogorov complexity and Levin complexity as the right measures for information. We make the folllowing observation: 

\begin{lemma}\label{L1}
If we can definie a computable bijection $p: \mathbb{N} \rightarrow \mathbb{N}$  for which in the limit the information efficiency is unbounded then computional processes generate information beyond the information stored in the program $p$ it self, i.e. equation \ref{E1} is invalid. Consequently equation \ref{E2} is the right bound. 
\end{lemma}
Proof: First observe that most natural numbers are typical, i.e. random. Their Kolmogorov complexity is ``close'' to the logarithm of the value. Since $p$ computes a bijection on $\mathbb{N}$, both input and output will contain a ``sufficient'' amount of random, i.e. incompressible, elements to the effect that the equations \ref{E1} and \ref{E2} describe equalities for dense sets at any scale. Now consider equation \ref{E1}, which can be rewritten as: 

\[K(Output) - K(Input) =\]
\begin{equation}\label{IE3} 
\delta(Program(Input)) \leq  K(Program)  + O(1) = c
\end{equation}

i.e. the information efficieny is bounded which contradicts the assumption. Now rewrite \ref{E2} as: 

\[K_t(Output) - K(Input) =\]
\begin{equation}\label{IE4} 
\delta(Program(Input)) \leq  K(Program) +\log(time(Program)) + O(1)
\end{equation}

with the extra factor $\log(time(Program))$.  This is  the right bound since, following observation \ref{O1}, $\log(time(Program))$  is the maximum speed at which deterministic computational processes generate information.   $\Box$

\section{Computable functions that generate and discard information}

In this paragraph we show that there are indeed finite halting programs that generate and discard an unbounded amount of information in the limit. Central is the notion of an \emph{elastic transformation} of the set $\mathbb{N}^2$. 

\subsection{The Cantor packing function}
Observe that there is a two-way polynomial time computable bijection $\pi: \mathbb{N}^2 \rightarrow \mathbb{N}$ in the form of the so-called Cantor packing function: 

\begin{figure}[ht!]
\centering
\includegraphics[width=90mm]{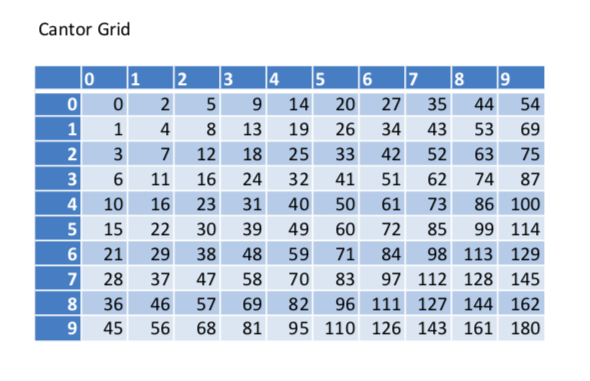}
\caption{An initial segment of the Cantor packing function \label{Cantor_Function}}

\end{figure}

\begin{equation}\label{CANTORPAIRING}
\pi(x,y) :=  \frac{1}{2}(x + y)(x + y + 1)+y
\end{equation}

An example of the computations involved in the bijection is given in the Appendix in paragraph \ref{A2}.  The Fueter - P\'{o}lya theorem \cite{FP23} states that the Cantor pairing function and its symmetric counterpart $\pi'(x,y)=\pi(y,x)$ are the only possible quadratic pairing functions.  A segment of this function is shown in figure \ref{Cantor_Function}. The information efficiency of this function is: 

\begin{equation}\label{CantEff}
\delta(\pi(x,y)) = \log (\frac{1}{2} (x+y+1) (x+y) + y) - \log x - \log y
\end{equation}

\begin{figure}[ht!]
\centering
\includegraphics[width=90mm]{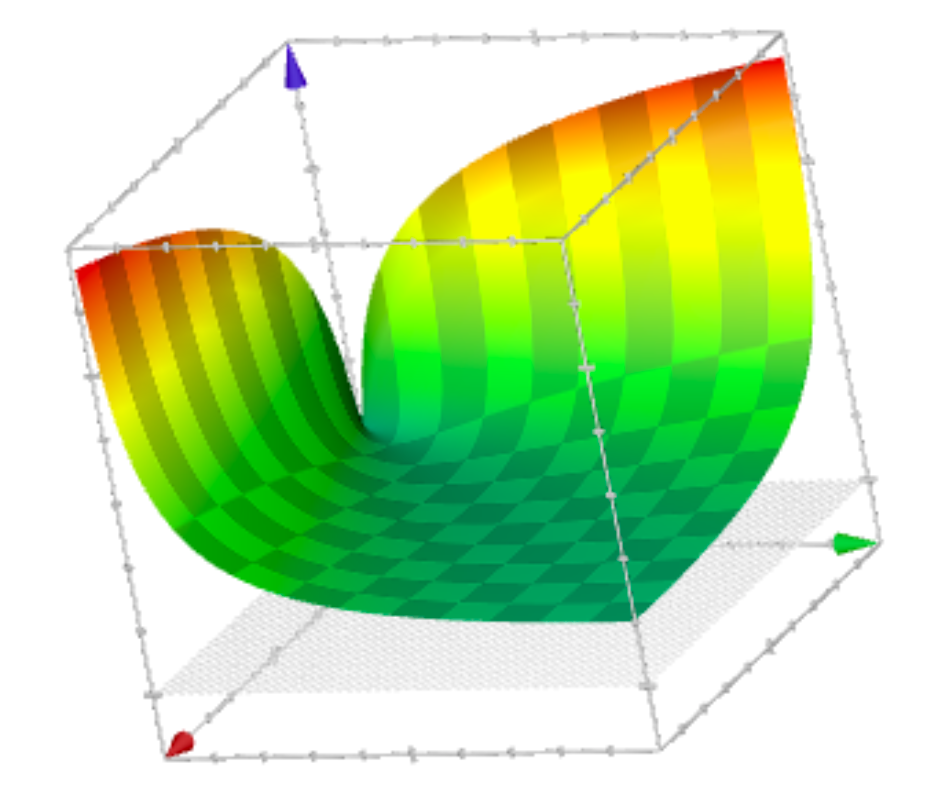}
\caption{The information efficiency of the Cantor packing function, $0< x < 10^9$, $0< y < 10^9$, $-1< z < 7$. The shaded area is the $z=0$  surface. \label{Cantor_Efficiency}}
\end{figure}

\begin{observation}\label{Fold1}
The Cantor function defines what one could call: a  \emph{discontinuous folding operation over the counter diagonals}. On the line $y=0$ we find the images $1/2x(x-1) =\Sigma_{i}^{x} i $. For points $(x_1,y_1)$ and $(x_2,y_2)$ on different counter diagonals  $x_1 + y_1= c_1$ and   $x_2 + y_2= c_2$ we have that $|\pi(x_1,y) - \pi(x_2,y)| = c$.  Equation \ref{CantEff} can be seen as the description of an information topology. The Cantor function runs over the counter diagonals and the image shows that the information efficiencies of points that are in the same neighborhood are also close. 
\end{observation}

\begin{observation}
The information efficiency of the Cantor packing function has \emph{infinite precision} (see Figure \ref{Cantor_Efficiency}). 
\end{observation}

This is what one would expect from a function that defines a polynomial time computable bijection.  We analyse some limits that define the information efficiency of the function. On the line $y=x$ we get: 

\begin{equation}\label{EffDiag}  
\lim_{x \rightarrow \infty} \delta \pi(x,x) = \lim_{x \rightarrow \infty}\log(\frac{1}{2} (2x + 1)(2x) ) - 2\log x =
\end{equation}

\[ \lim_{x \rightarrow \infty}\log \frac{ 2x^2 + x }{x^2} =  1 \] 

For the majority of the points in the space $\mathbb{N}^2$ the function $\pi$ has an information efficiency close to one bit. On every line through the origin $y=hx$ $(h>0)$ the information efficiency in the limit is constant: 

\begin{equation}\label{EffOrig}
\lim_{x \rightarrow \infty}  \delta(\pi(x,hx)) = 
\end{equation}
\[  \lim_{x \rightarrow \infty}  \log (\frac{1}{2} (x+hx+1) (x+hx) + hx) - \log x - \log hx = \]  
\[\log(1/2(h+1)^2) - \log h \]

Yet on every line $y=c$ (and by symmetry $x=c$) the information efficiency is unbounded: 

\begin{equation}\label{EffConst}
 \lim_{x \rightarrow \infty}  \delta(\pi(x,c)) = 
\end{equation}
\[  \lim_{x \rightarrow \infty}  \log (\frac{1}{2} (x+c+1) (x+c) + c) - \log x - \log c =  \infty \]  

Together the equations \ref{CANTORPAIRING}, \ref{CantEff}, \ref{EffDiag}, \ref{EffOrig} and  \ref{EffConst} characterize  the basic behavior of the information efficiency of the Cantor function.

\begin{figure}[ht!]
\centering
\includegraphics[width=120mm]{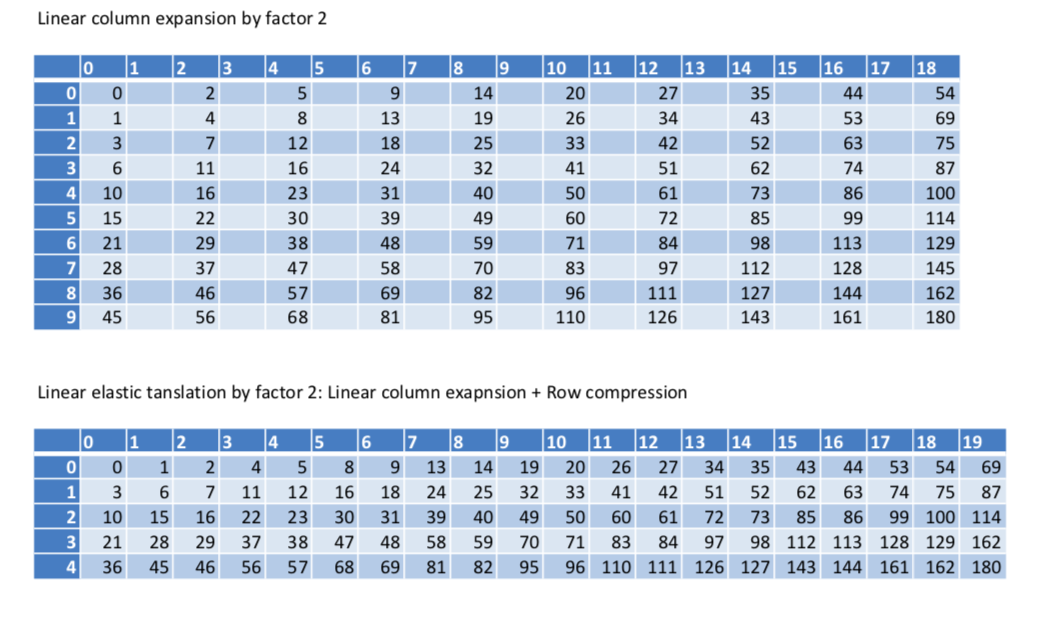}
\caption{Row expansion by  factor $2$ and elastic translation by a factor 2 for the segment in Figure  \ref{Cantor_Function}. For a computation of the exact information efficiency see figure \ref{ThreeDInformation_efficiency}.  \label{Cantor_Elasticity}}
\end{figure}

\begin{figure}[ht!]
\centering
\includegraphics[width=90mm]{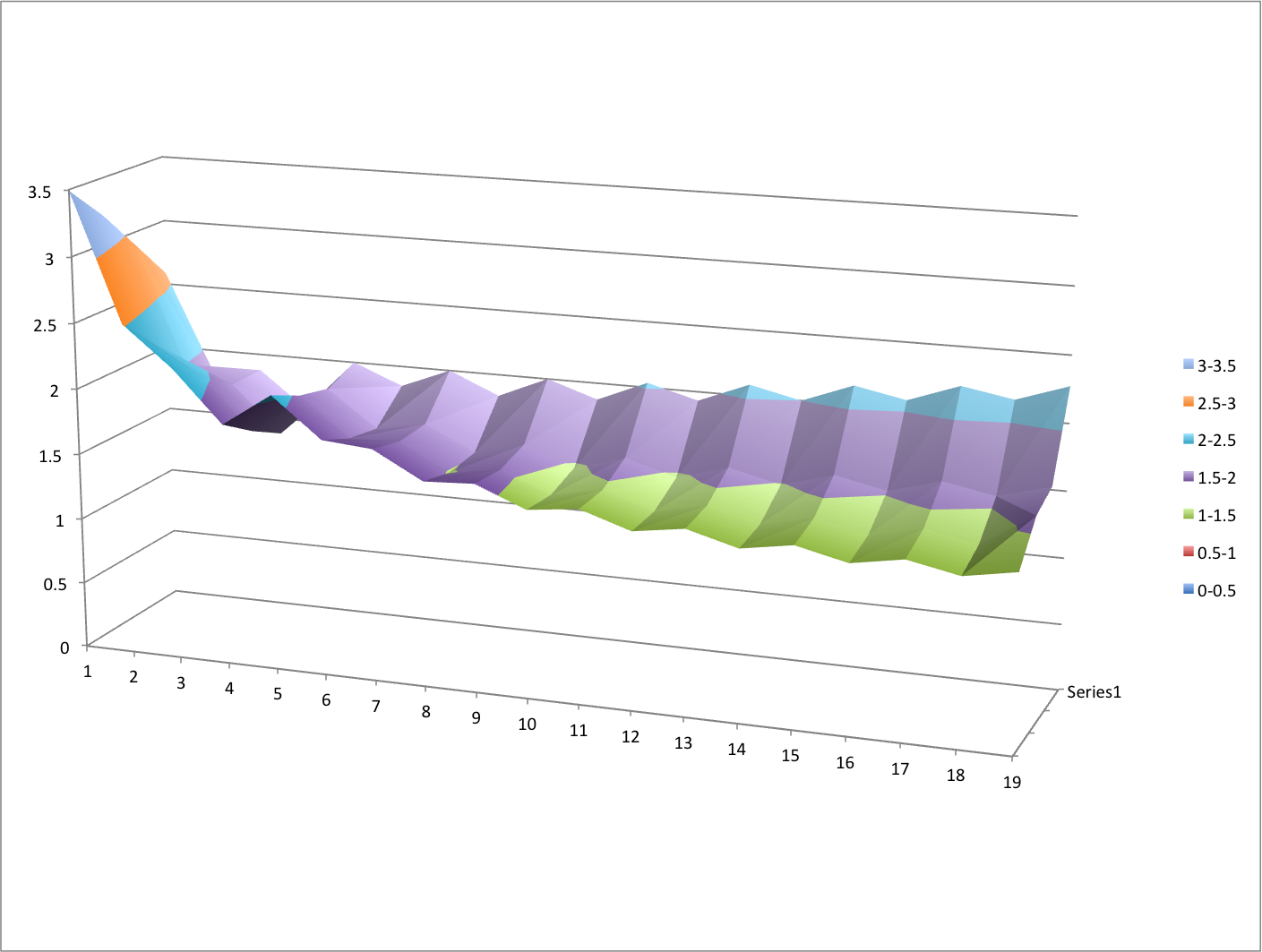}
\caption{An exact computation of the point by point information efficiency after an elastic transformation by a factor $2$  for a segment (1-4 by 1-19) of the table in figure \ref{Cantor_Elasticity}. The existence of two seperate interleaving information efficiency functions is clearly visible in line with definition \ref{InfEff}.  \label{ThreeDInformation_efficiency}}
\end{figure}

\begin{figure}[ht!]
\centering
\includegraphics[width=90mm]{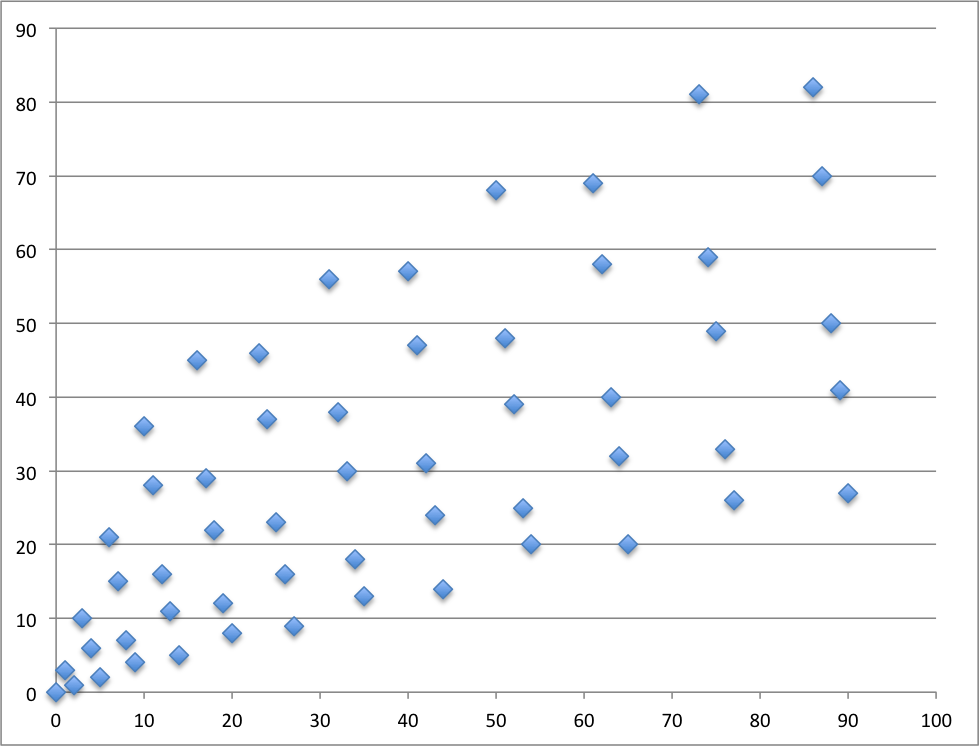}
\caption{A fragment of the bijection generated by an elastic transformation by a factor $2$  for a segment (0-4 by 0-9) of the table in figure \ref{Cantor_Elasticity}. The bijection $\pi(\epsilon_2(\pi^{-1})): \mathbb{N} \rightarrow \mathbb{N}$ generates a cloud of compressible and expandable points. The compressible points are above the line $x=y$.  \label{bijection_shift2}}
\end{figure}

\subsection{Linear elastic transformations of the Cantor space}

 Actually equation  \ref{EffConst}  is responsible for remarkable behavior of the Cantor function under what one could call \emph{elastic transformations}. For elastic transformations we can compress the Cantor space along the $y$-axis by any constant without actually losing information. Visually one can inspect this counter intuitive phenomenon in figure \ref{Cantor_Efficiency} by observing the concave shape of the information efficiency function: at the edges ($x=0$, $y=0$) it has in the limit an unbounded amount of compressible information. The source of this compressibility in the set $\mathbb{N}$ is the set of numbers that is logarithmically close to sets $\frac{1}{2}x(x+1)$ and  $\frac{1}{2}y(y+1) + y)$. In terms of Kolmogorov complexity these sets of points define regular dips of depth $\frac{1}{2} \log x$ in the integer complexity function that in the limit provide an infinite source of highly compressible numbers.  In fact, when we would draw  figure \ref{Cantor_Efficiency} at any scale over all functions $f: \mathbb{N}^2 \rightarrow \mathbb{N}$ we would see a surface with all kinds of regular and irregular elevations related to the integer complexity function.

 Observe figure  \ref{Cantor_Elasticity}. The upper part shows a discrete translation over the $x$-axis by a factor $2$. This is an information expanding operation: we add the factor $2$, i.e. one bit of information, to each $x$ coordinate. Since we expand information, the density of the resulting set in $\mathbb{N}^2$ also changes by a factor 2. In the lower part we have distributed the values in the columns $0,2,4,\dots$ over the  columns $0-1,2-3,4-5,\dots$. We call this an elastic translation by a factor $2$. The space $10 \times 10 = 100 $ is transformed in to a $5 \times 20 = 100$  space. The exact form of the translation is: $\epsilon_2(x,y) = (2x + (y\bmod{2}), \lfloor \frac{y}{2} \rfloor)$.

\begin{figure}[ht!]
\centering
\includegraphics[width=90mm]{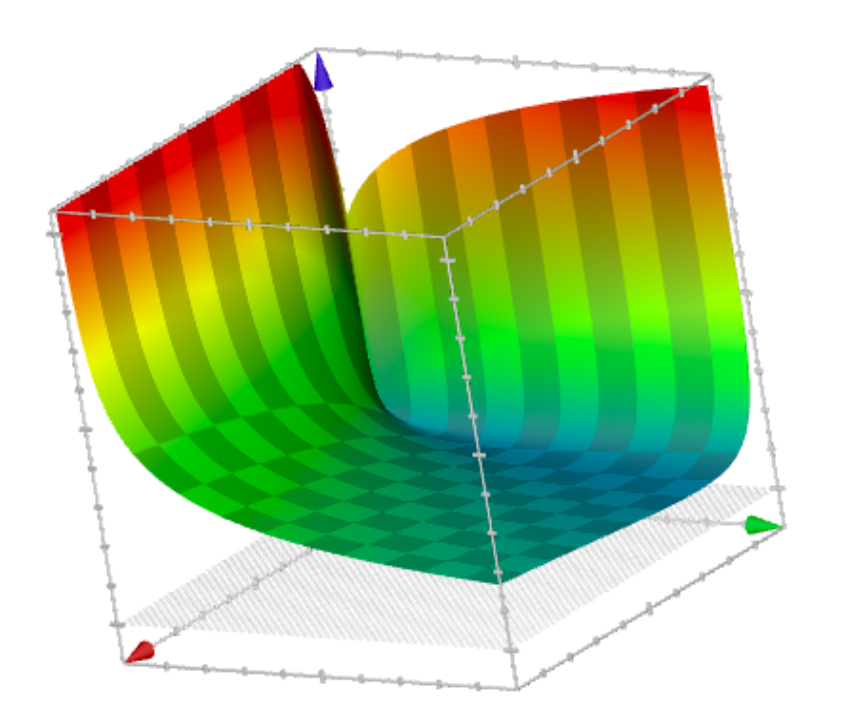}
\caption{The information efficiency of the reference function of the Cantor packing function on the same area as in figure \ref{Cantor_Efficiency}  after an elastic shift by a factor 100, $0< x < 10^9$, $0< y < 10^9$, $-1< z < 10.6$. The shaded area is the $z=0$  surface. \label{Elastic2}}
\end{figure}

 The effect of this elastic translation on the information efficiency on a local scale can be seen in figure \ref{ThreeDInformation_efficiency}.  After some erratic behavior close to the origin the effect of the translation evens out. There are traces of a phase transition: close to the origin the size of the $x$, $y$ coordinates is comparable to the size of the shift $c$, which influences the information efficiency considerably. From the wave pattern in the image it is clear that a linear elastic transformation by a factor $c$ essentially behaves like a set of $c$ functions (in this case $2$), each with a markedly different information efficiency.

Even more interesting is the behavior, shown in figure \ref{bijection_shift2}, of the bijection:  
 
 \begin{equation}\label{B1}
\begin{tikzcd}
\mathbb{N}\arrow{r}{\pi^{-1}}   & \mathbb{N}^2\arrow{r}{\epsilon_2} & \mathbb{N}^2\arrow{r}{\pi} & \mathbb{N}
\end{tikzcd} 
\end{equation}

Although the functions $\pi$, $\pi^{-1}$ and $\epsilon_2$ are bijections and can be computed point wise in polynomial time, all correlations between the sets of numbers seems to have been lost. The reverse part of the bijection shown in formula \ref{B2} seems hard to compute, without computing large parts of \ref{B1} first. 
 
 \begin{equation}\label{B2}
 \begin{tikzcd}
\mathbb{N}\arrow{r}{\pi^{-1}}   & \mathbb{N}^2\arrow{r}{\epsilon_2^{-1}} & \mathbb{N}^2\arrow{r}{\pi} & \mathbb{N}
\end{tikzcd} 
\end{equation}

 \begin{observation}\label{Fold2}
Linear elastic transformations introduce a second type of \emph{horizontal discontinuous folding operations over the  columns.}. These operations locally distort the smooth topology of the Cantor function into clouds of isolated points.
\end{observation}
 
 On a larger scale visible in figure \ref{Elastic2} we get a smooth surface. The distortion of the symmetry compared to figure \ref{Cantor_Efficiency} is clearly visible.  In accordance with equation  \ref{EffConst}, nowhere in the set $\mathbb{N}^2$ the information efficiency is negative. In fact the information efficiency is lifted over almost the whole surface. On the line $x=y$ the value in the limit is: 

\begin{equation}\label{EffDiag2}  
\lim_{x \rightarrow \infty} \delta \pi(\epsilon_2((x,x))) =
\end{equation}
\[ \lim_{x \rightarrow \infty} \log (\frac{1}{2} (2x+x/2+1) (2x+x/2) + x/2) - \log 2x - \log x/2 = 2 \log 5 - 3\]

\begin{figure}[ht!]
\centering
\includegraphics[width=90mm]{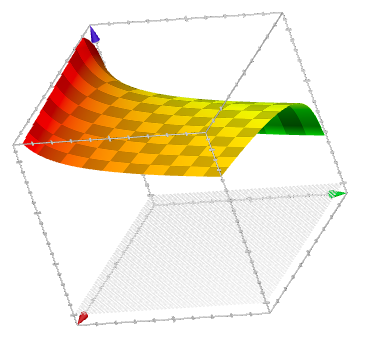}
\caption{The first information efficiency function of the Cantor packing function on the same area as in Figure \ref{Cantor_Efficiency}  after an elastic shift by a factor 100, $0< x < 10^9$, $0< y < 10^9$, $-1< z < 17$. The shaded area is the $z=0$  surface. For this transformation we have  $100$ different information efficiency functions. Only the first is shown.  \label{Elastic_Shift100}}
\end{figure}

A more extreme form of such a distortion can be seen in figure \ref{Elastic_Shift100} that shows the effect on the information efficiency after an elastic translation by factor $100$. Clearly the lift in information efficiency over the whole surface can be seen. We only show the first of $100$ different information efficiency functions here. Computed on a point by point basis we would see periodic saw-tooth fluctuations over the  $x$-axis with a period of $100$. This discussion shows that elastic transformations of the Cantor space act as a kind of \emph{perpetuum mobile} of information creation. For every elastic transformation by a constant $c$ the information efficiency in the limit is still positive:  

\begin{lemma}\label{Crux}
No compression by a constant factor $c$ along the $y$-axis (or $x$-axis, by symmetry) will generate a negative information efficiency in the limit. 
\end{lemma}
Proof: immediate consequence of equation  \ref{EffConst}. The information efficiency is unbounded in the limit on every line $x=c$ or $y=c$. $\Box$ 

\subsection{A general model of elastic transformations} 

In the following we will study, what we call \emph{general elastic transformations} of the space $\mathbb{N}^2$:

\begin{definition}\label{DEfElTrans}
The function $\epsilon_f: \mathbb{N}^2 \rightarrow \mathbb{N}^2$ defines an \emph{elastic translation} by a function $f$ of the form: 

\begin{equation}\label{GeneralShift}
\epsilon_{r}(x,y) = (xf(x) + (y\bmod{r(x)}), \lfloor \frac{y}{r(x)} \rfloor)
\end{equation}

Such a transformation is \emph{super-elastic} when: 
\[\lim_{x \rightarrow \infty} r(x) = \infty\]
It is \emph{polynomial} when it preserves information about $x$ : 
\[\lim_{x \rightarrow \infty} r(x) = cx^k\]
It is \emph{linear} when:
 \[r(x)=c\]  
 The \emph{reference function} of the translation is: 
\[\epsilon'_{r}(x,y) = (r(x)x, \frac{y}{r(x)})\]  

  We will assume that the function $f$ can be computed in time polynomial to the length of the input. 
\end{definition}

Observe that the reference function: $\epsilon'_{r}(x,y) = (r(x)x, \frac{y}{r(x)})$ is information neutral on the arguments: 

\[\log x + \log y - (\log r(x)x + \log \frac{y}{r(x)})  = 0\]

An elastic translation consists from an algorithmic point of view of two additional operations: 
\begin{enumerate}
\item An \emph{information discarding} operation on $y: \frac{y}{r(x)} \rightarrow  \lfloor \frac{y}{r(x)} \rfloor$. 
\item An \emph{information generating} operation on $x: r(x)x \rightarrow r(x)x + (y\bmod{r(x)})$.
\end{enumerate} 

\begin{figure}[ht!]
\centering
\includegraphics[width=90mm]{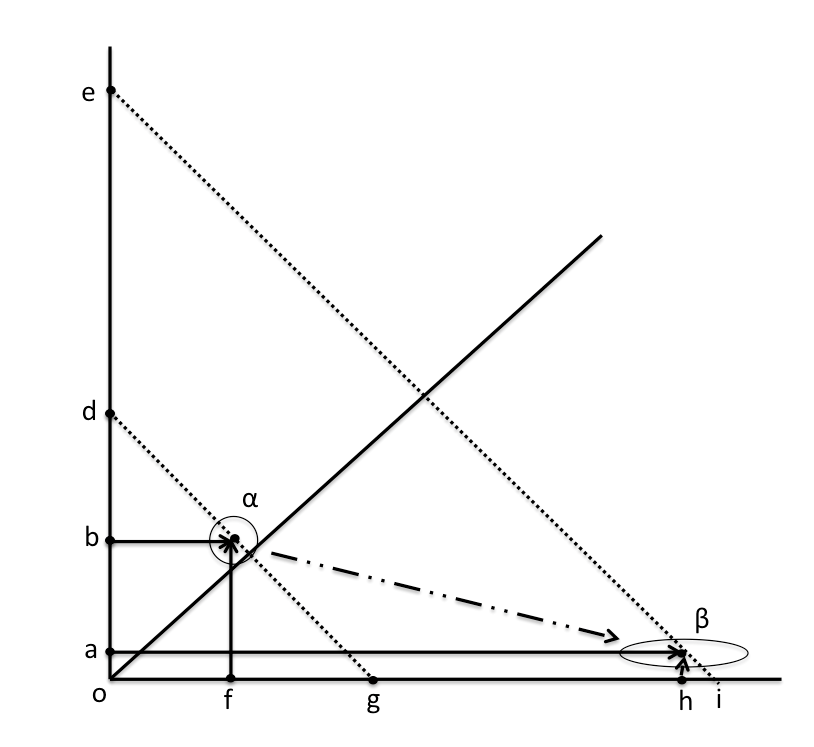}
\caption{An elastic translation from neighborhood $\alpha$ to neighborhood $\beta$.  \label{Cantor_Shift}}
\end{figure}

 A schematic overview of a linear elastic shift is given in figure \ref{Cantor_Shift}. Here the letters $a,\dots i$ are natural numbers. An arbitrary point  in neighborhood  $\alpha$ with coordinates $(b,f)$ close to the diagonal is translated to point $(a,h)$  in neighborhood $\beta$. The formula for the translation is given by definition \ref{DEfElTrans}: $ \epsilon_r(x,y) = (r(x)x + (y\bmod{r(x)}), \lfloor \frac{y}{r(x)} \rfloor)$.  We have $a = \lfloor \frac{b}{r(f)} \rfloor$ and $h = r(f)e + (b\bmod{r(f)})$.  
 
 The information efficiency of  an \emph{elastic transformation} is: 
 
\[\delta(\pi(\epsilon_{r})(x,y)) =  \delta(\pi( r(x)x + (y\bmod{r(x)}),  \lfloor \frac{y}{r(x)} \rfloor))\]

 The information efficiency of  a \emph{linear elastic transformation} by a factor  $f(x)=c$  is: 
 
 \[\delta(\pi( cx + (y\bmod c ),  \lfloor \frac{y}{c} \rfloor))\]

\begin{observation}
Elastic transformations by a constant $c$ of the Cantor space replace the highly efficient Cantor packing function with $c$ different interleaving functions, each with a different information efficiency.  Equation \ref{GeneralShift} must be seen as a \emph{meta-function} or \emph{meta-program} that spawns off $c$ different new programs. 
\end{observation}

This is illustrated by the following lemma:  

 \begin{lemma}\label{L2}
Linear elastic translations generate information. 
\end{lemma}
Proof: This is an immediate effect of the use of the mod function. A \emph{linear elastic translation by a constant $c$} of the form: \[\epsilon_{c}(x,y) = (cx + (y\bmod c), \lfloor \frac{y}{c} \rfloor)\] has $c$ different information efficiency functions. Note that the function $x \bmod c$ produces all numbers $d < c$, including the incompressible ones that have no mutual information with $c$: $K(d|c)= \log d + O(1)$.For each value  $d = y\bmod c$ we get a function with different information efficieny: 
\[\delta(\pi(\epsilon_{c,d}(x,y)))) = \log \pi (cx + d, \lfloor \frac{y}{c} \rfloor) -  \log x - \log y\]

On a point by point basis the number $d$ is part of the information computed by $\pi(\epsilon_{c,d}(x,y)))$. In other words the computation $\pi(\epsilon_{c,d}(x,y))$ adds information to the input for specific pairs $(x,y)$ that is not available in the formula for $\epsilon_{c}(x,y)$.  The effect is for linear transformations constant in the limit, so it is below the accuracy of Kolmogorov complexity.
$\Box$

For typical cells $(x,hx)$ on a line $y=hx$ the function  $\delta(\pi(\epsilon'_{c}(x,y)))$ gives in the limit a constant shift which can be computed as: 

  \begin{equation}\label{E5} 
\lim_{x \rightarrow \infty}\frac{\pi(\epsilon'_{c}(x,y))}{\pi(x,y)}   =
\lim_{x \rightarrow \infty}\frac{ \frac{1}{2} (c x + \frac{h x}{c} + 1) (c x +  \frac{h x}{c}) +  \frac{h x}{c}}{ \frac{1}{2} (x + h x + 1) (x + h x) + h x} = \frac{(c^2 + h)^2}{c^2(1 + h)^2} \geq 1
\end{equation} 

Note that this value is only dependent on $h$ and $c$ for all $c \in \mathbb{N}^{+}$ and $r \in \mathbb{R}^{+}$. The general lift of the line $x=y$ for an elastic shift by a constant $c$ is: 

\begin{equation}\label{ConShift} 
 \lim_{x \rightarrow \infty} \log (\frac{1}{2} (cx+\frac{x}{c}+1) (cx+\frac{x}{c}) + \frac{x}{c})) - \log cx - \log \frac{x}{c} =
\end{equation}

\[- \log \frac{2}{c} - \log c +\log (2 + \frac{1}{c^2} + c^2)\]

 We get a better understanding of the extreme behavior of the reference function $\delta(\pi(\epsilon'_{c}(x,y)))$ when we rewrite equation \ref{E5} as: 

  \[ \frac{(c^2 + h)^2}{c^2(1 + h)^2} = \frac{c^2} { 1 + 2h + h^2 }+  \frac{ 2h} {  1 + 2h + h^2 } +  \frac{ h^2}{ c^2 + 2c^2h + c^2h^2 } \geq 1\]
 
and take the following limit: 
  
     \[ \lim_{h \rightarrow \infty} \frac{(c^2 + h)^2}{c^2(1 + h)^2} =  c^{-2} \]  

If $c$ is constant then it has small effects on large $h$ in the limit. The reference function allows us to study the dynamics of well-behaved ``guide points'' independent of the local distortions generated by the information compression and expansion operations. Note that elastic transformations start to generate unbounded amounts of information in each direction $y=hx$ in the limit on the basis of  equation \ref{E5}:

  \begin{equation}\label{E6} 
 \lim_{c \rightarrow \infty} \frac{(c^2 + h)^2}{c^2(1 + h)^2} = \infty
\end{equation} 
   
If $c$ grows unboundedly then the information efficiency of the corresponding reference functions goes to infinity for every value of $h$. Consequently,  when $c$ goes to infinity, the reference functions predict infinite information efficiency in $\mathbb{N}^2$ in all directions, i.e. we get infinite expansion of information in all regions without the existence of regions with information compression. This clearly contradicts central results of Komogorov complexity if we asume that elastic translations are defined in terms of a single program. The situation is clarified by the proof of lemma \ref{L2}: if $c$ goes to infinity we create an unbounded amount of new functions that generate an unbounded amount of information.

\subsection{Polynomial  transformations}    
The picture that emerges from the previous paragraphs is the following: we can define bijections on the set of natural numbers that generate information for almost all numbers. The mechanism involves the manipulation of clouds of points of the set $\mathbb{N}$: sets with density close to the origin are projected into sparse sets of points further removed from the origin. This proces can continue indefinitely.

In this context we analyse polynomial translations. The simplest example is the elastic translation by the factor $x$:

\begin{figure}[ht!]
\centering
\includegraphics[width=120mm]{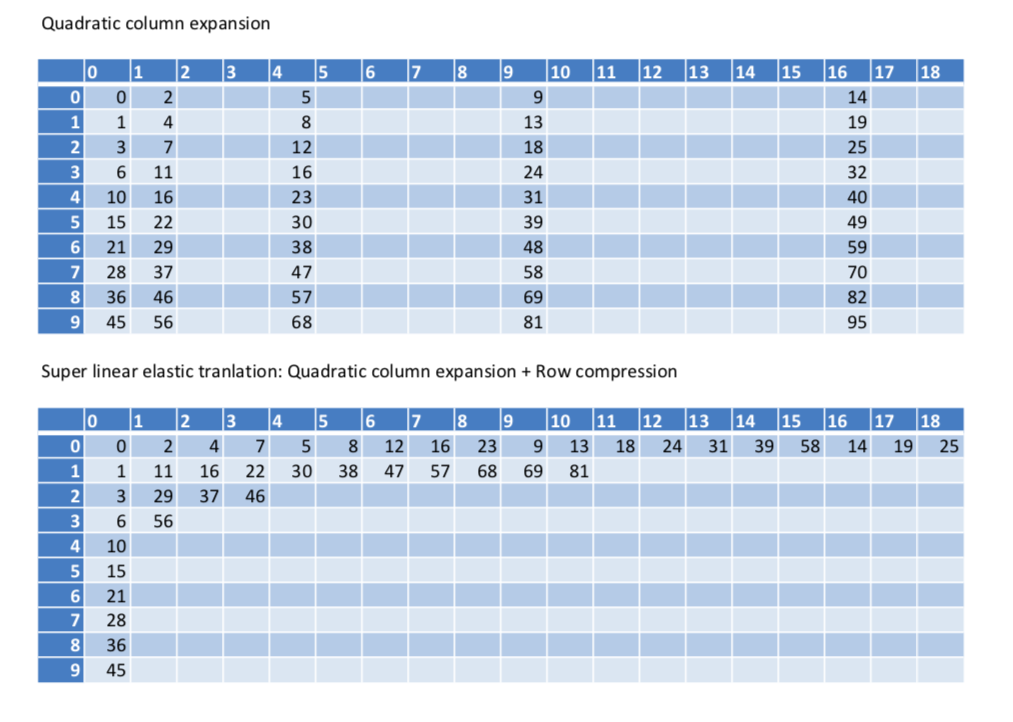}
\caption{Row expansion by a superlinear, $r(x)=x$ factor and a fragment of the corresponding  elastic translation for the segment in Figure  \ref{Cantor_Function} \label{Cantor_Super_Elasticity}}
\end{figure}

\begin{equation}\label{SupLin}
\epsilon_x(x,y) = (x^2 + (y\bmod{x}), \lfloor \frac{y}{x} \rfloor)
\end{equation}

\begin{figure}[ht!]
\centering
\includegraphics[width=120mm]{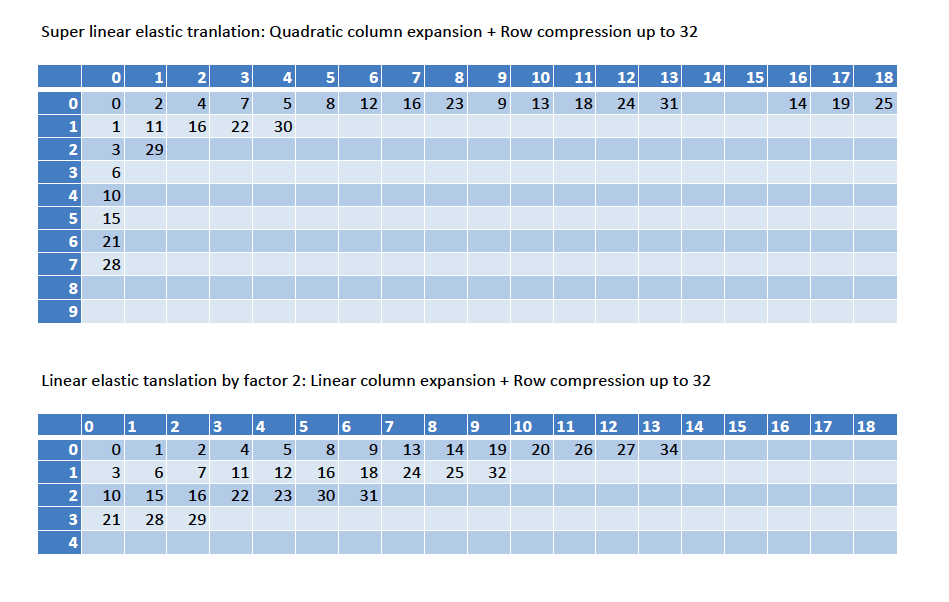}
\caption{Above: Super linear expansion by a function $r(x)=x$ up to 32.   Below: Linear elastic expansion by a factor $2$ up to the number $32$.  \label{Discontinuous}}
\end{figure}

A tiny fragment of the effects is shown in figure \ref{Cantor_Super_Elasticity}. There does not seem to be a fundamental difference compared to the previous examples and there seems to be no difficulty in constructing such a translation along the lines suggested in the figure. However upon closer inspection things  are different as can be observed in figure \ref{Discontinuous}. The first table shows the computation of the function up to the number $32$. Observe that the columns $14$ and $15$ are empty, while column $16-18$ have a value. This effect does not appear in the second table. Linear elastic translations induce a change in the direction of the iso-information line $(d,g)$, but the image of the translation keeps a coherent topology at any stage of the computation.  

Polynomial translations on the other side are discontinuous. They tear the space apart in to separate regions.  An appropriate metaphor would be the following: expansion away from the origin over the $x$-axis, sucks a \emph{vacuum} that must be filled by a contraction over the $y$-axis. Actually the creation of such a vacuum is an information discarding operation. The analysis above shows that the vacuum created by the shift described by equation \ref{SupLin} for cells on the line $y=0$ is bigger than the \emph{whole surface} of the triangle $(o,d,g)$. The effect is that the image of the translation becomes discontinuous.

\begin{theorem}\label{T1}
Polynomial elastic transformations of $\mathbb{N}^2$: 
\begin{enumerate}
\item Discard  and expand  information on the line $y=0$ unboundedly. 
\item  Project a dense part of $\mathbb{N}^2$ and  $\mathbb{N}$ on $y=0$.
\item Generate an unbounded amount of information for typical points in $\mathbb{N}^2$ in the limit.   
\end{enumerate}
\end{theorem}
Proof: The formula for a polynomial translation is: 

\[\epsilon_{r}(x,y) = (cx^{k+1} + (y\bmod{cx}), \lfloor \frac{y}{cx} \rfloor)\]

Take $k= c=1$. Consider a typical point  $f$ in figure \ref{Cantor_Shift}. We may assume that the numbers $d$ and $g$ are typical (i.e. incompressible and thus $\pi(g,d)$ is inompressible too.   Remember that the Cantor function runs over the counter diagonal which makes the line $(d,g)$ an \emph{iso-information} line. 
\begin{enumerate}
\item Discard  and expand  information on the line $y=0$ unboundedly: 
\begin{itemize}
	\item Discard  information: Horizontally point $(f,0)$ will be shifted to location $(f^2,0)$ and $(f(+1),0)$ to $(f^2 + 2f + 1, 0)$. The cantor index for point $(f,0)$ is $\frac{1}{2}f(f-1) < f^2$. 
	\item Expand information: The cells from $(f,0)$ to $f(f,2f)$ will be ``padded'' in the strip between  $(f^2,0)$ and $(f^2 + 2f + 1, 0)$. But this operation ``steals'' a number of  $f$ cells from the domain above the line $(d,g)$. Now take a typical point $n$ on location  $(f,m)$. such that $f < m < 2f$. This point will land at  $(f^2 + m,0)$ somewhere between $(f^2,0)$ and $(f^2 + 2f + 1, 0)$, which gives: 
\[\pi(f,(f+m))= \frac{1}{2} (f + (f +m) +1)(f +(f+m)) + (f+m) = \]
\[ 2 f^2 + \frac{1}{2}fm + m^2 +  3f + 1\frac{1}{2} (m + fm)   \]
\[  \gg f^2 + 2f + 1 \]
\end{itemize}
 Note that the effects are dependent on $f$ and $m$, so they are unbounded in the limit.  Alternatively observe the fact that polynomial transformations are superelastic and apply equation \ref{E6}.
\item  Project a dense part of $\mathbb{N}^2$ and  $\mathbb{N}$ on $y=0$: For every point $f$ all the cells  $(f,0)$ to $f(f,2f)$  up to the line $y = 2x$ will end up on line $y=0$. since all points in this dense region are projected on the line $y=0$ most points $\pi(f,(f+m))$ are incompressible. 
\item Generate an unbounded amount of information for typical points in $\mathbb{N}^2$ in the limit. Take a typical point $(x,y)$ such that  $\log x \approx \log y$:

\begin{equation}\label{INFGROW}
\pi(\epsilon_{r}(x,y)) >  \pi(\epsilon_{r}(x,1)) = \frac{1}{2}( x^2+2)(x^2+1) +1 > \frac{1}{2}x^4
\end{equation}
\[\delta (\pi(\epsilon_{r}(x,y))) >  \log  \frac{1}{2}x^4  - \log x - \log y \approx 4 \log x - 2 \log x = 2 \log x \]

which gives $K\pi(\epsilon_{r}(x,y)) > K(\pi(x,y)) + 2 \log x$. 
\end{enumerate}
This argument can easily be generalized to other values of $c$ and $k$. $\Box$ 

Polynomial shifts generate information above the asymptotic sensitivity level of Kolmogorov complexity.  Note that $\epsilon_{r}(x,y)$ is still  a computable bijection:  

 \[\forall (x,y)_{\in \mathbb{N}^2} \exists (u,v)_{\in \mathbb{N}^2} (\epsilon_{r}(u,v)= (x,y))\]   
 
 \[u^2 \leq x < (u+1)^2 \]  
 \[a = (u+1)^2 - u^2\]
 \[y = \lfloor \frac{v}{a} \rfloor \]
 
 This analysis holds for all values $\delta (\pi(\epsilon_{r}(x,y))) \in \mathbb{N}$ including the values that are typical, i.e. incompressible. 

\begin{observation}An immediate consequence is the translation $\epsilon_{r}(x,y) = (cx^{k+1} + (y\bmod{cx}), \lfloor \frac{y}{cx} \rfloor)$ must be interpreted as a \emph{function scheme}, that produces a countable set of new functions. One for each column $x=c$. Actually $x$ can be seen as an \emph{index} of the function that is used to compute $\epsilon_{r}(u,v)= (x,y)$.
\end{observation}

The difference between linear and polynomial elastic transformations marks the phase transition between continuous and non-continuous shifts. When we use the enumeration of the Cantor packing function to compute such shifts the time needed to compute the location of certain points on the line $y=0$ close to each other may vary exponentially in the representation of the numbers involved. On the other hand such translations can be easily computed as bijections on a point by point basis in polynomial time, since the $x$ coordinate of the image codes the index of the function that was used to compute it. 

\subsection{The information efficiency of arithmetical functions on sets of numbers}

In this paragraph we study elastic translations that grow faster than any polynomial function based on elementary arithmetical functions. These translations are so ``aggressive''  that they destroy the topology of the space completely and project chaotic clouds of points on the $x$-axis A crucial tool for the construction of super polynomial transformations is the bijective mapping of finite sets of numbers in $\mathbb{N}$ to the space  $\mathbb{N}^2$:

\begin{definition}

$\mathcal{P}(\mathbb{N})$ is the powerset or set of subsets of $\mathbb{N}$. $\mathfrak{P}(\mathbb{N})$  the set of finite subsets of $\mathbb{N}$. A \emph{characteristic function} of an infinite subset of $S \subseteq \mathbb{N}$ is $f_S: \mathbb{N} \rightarrow \mathbb{N}$, a monotone ascending function such that $\forall(x) f_S(x+ 1) > f(x)$. Here $x$ is the \emph{index} of $f_S(x)$ in $S$. 
 \end{definition}
 
$\mathcal{P}(\mathbb{N})$ is uncountable, whereas $\mathfrak{P}(\mathbb{N})$ can be counted. Consequently $\mathcal{P}(\mathbb{N})  - \mathfrak{P}(\mathbb{N})$ is also uncountable. Proofs of the countability of $\mathfrak{P}(\mathbb{N})$ rely on the axiom of choice to distribute set  $\mathfrak{P}(\mathbb{N})$ in to partitions with the same cardinality. A useful concept in this context is the notion of \emph{combinatorial number systems}: 

\begin{definition}\label{COMBNUMBSYS}
The function $\sigma_k:\mathbb{N}^k \rightarrow \mathbb{N}$ defines for each element \[s = (s_k,\dots,s_2,s_1) \in \mathbb{N}^k\]  with the strict ordering $s_k > \dots s_2 > s_1 \geq 0$ its index in a $k$-dimensional \emph{combinatorial number system} as:

\begin{equation}\label{COMBSYST}
\sigma_k(s)= {s_k \choose k} + \dots + {s_2 \choose 2} + {s_1 \choose 1}
\end{equation}
\end{definition}

The function $\sigma_k$ defines for each set $s$ its index in the lexicographic ordering of all sets of numbers with the same cardinality $k$. The correspondence does not depend on the size $n$ of the set that the $k$-combinations are taken from, so it can be interpreted as a map from $\mathbb{N}$ to the $k$-combinations taken from $\mathbb{N}$. For singleton sets we have: $\sigma_1(x)= {x \choose 1}= x$, $x \geq 0$. For sets with cardinality $2$ we have:

\[\sigma_k(0)=  {1 \choose 2} + {0 \choose 1} \rightarrow  \{1,0\}\]
\[\sigma_k(1)=  {2 \choose 2} + {0 \choose 1} \rightarrow  \{2,0\}\]
\[\sigma_k(2)=  {2 \choose 2} + {1 \choose 1} \rightarrow  \{2,1\}\]
\[\sigma_k(3)=  {3 \choose 2} + {0 \choose 1} \rightarrow  \{3,0\} \dots\]

We can use the notion of combinatorial number systems to prove the following result: 

\begin{theorem}\label{COUNTCARD}
There is a bijection $ \phi:  \mathfrak{P}(\mathbb{N}) \rightarrow  \mathbb{N}$ that can be computed  efficiently. 
\end{theorem}
Proof: We prove the lemma for the set $\mathbb{N}^+$.  Let $S_k$ be the subset of all elements $s \in \mathfrak{P}(\mathbb{N}^+)$ with cardinality $|s| = k$. For each $k \in \mathbb{N}$ by definition \ref{COMBNUMBSYS} the set $S_k$ is described by a combinatorial number system of degree $k$. The function $\sigma_k:\mathbb{N}^{+k} \rightarrow \mathbb{N}$ defines for each element $s = (s_k,\dots,s_2,s_1) \in \mathbb{N}^{+k}$  with the strict ordering $s_k > \dots s_2 > s_1 \geq 0$ its index in a $k$-dimensional combinatorial number system. By definition \ref{COMBNUMBSYS} the correspondence is a polynomial time computable bijection. Now define $\phi_{||}^+: \mathfrak{P}(\mathbb{N}) \rightarrow \mathbb{N}$ as: 

\begin{equation}\label{CARD}
\phi_{||}^+(s) =   \pi((|s| - 1),\sigma_{|s|}(s) -1)
\end{equation}
Note that  both $\pi$ and $\sigma$ are computable bijections. When we have the set $s$ we can compute its cardinality $|s|$  in linear time and compute $\sigma_{|s|}(s)$ from $s$ in polynomial time. When we have $\phi_{||}^+(s)$ we can compute $|s|$ and compute $s$ from $\sigma_{|s|}(s)$ in polynomial time. 
$\Box$ 

\begin{figure}[!t]
\centering
\fbox{\includegraphics[ width=4in]{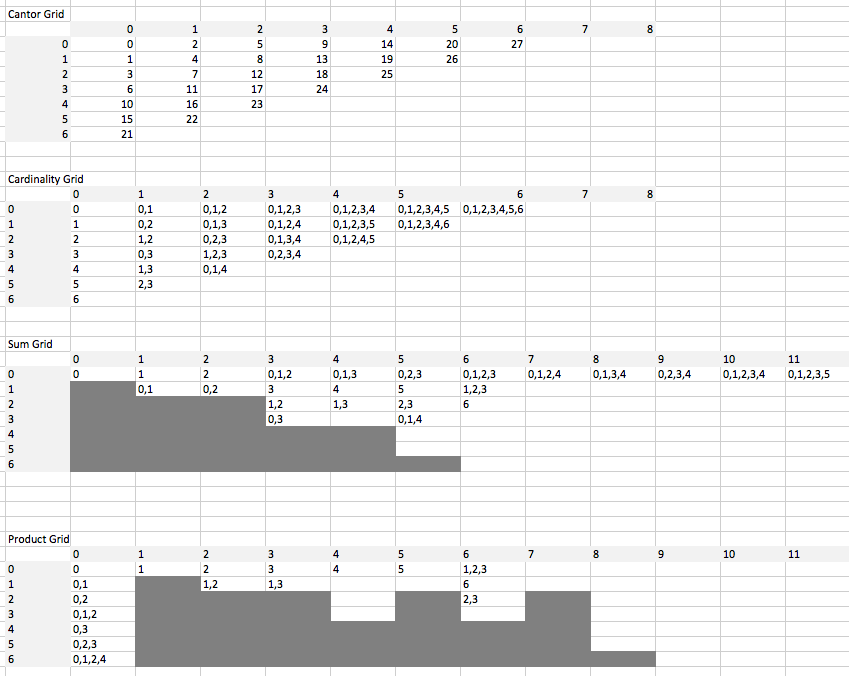}}
\caption{This figure illustrates the effects of two specific injections (based on sum and product)  of the form $\phi_{\zeta}(s) = \pi(\zeta (s),\theta_{\zeta(s)}(s))$ (equation \ref{GEN}). First table: basic Cantor grid. Second table: Mapping to finite sets of numbers. Third table: effect of the sum translation. Fourth table: effect of the product translation. }\label{GRIDEXAMPLES}
\end{figure}

An elaborate example of the computation both ways is given in the appendix in paragraph \ref{A2}. An example of the mapping can be seen in figure \ref{GRIDEXAMPLES} under the header Cardinality Grid. 

Note that in this proof the combinatorial number systems are defined on $\mathbb{N}^+$, while $\pi$ is defined on $\mathbb{N}^2$. The results can easily be normalized by linear time computable translations:  $s^+ = (s_k,\dots,s_2,s_1) \in \mathbb{N}^{+k} \rightarrow  s = (s_k -1,\dots,s_2 -1 ,s_1 -1 ) \in \mathbb{N}^{k}$.  For reasons of clarity, in the rest of the paper, we will ignore such corrections for $\mathbb{N}^+$ and use the function: 

\begin{equation}\label{CARD1}
\phi_{||}(s) =   \pi((|s| ),\sigma_{|s|}(s))
\end{equation}

The construction of the proof of theorem \ref{COUNTCARD} separates the set $\mathfrak{P}(\mathbb{N})$ in an infinite number of infinite countable partitions ordered in two dimensions: in the columns we find elements with the same cardinality, in the rows we have the elements with the same index.  

We give some examples. Consider the four tables in figure \ref{GRIDEXAMPLES} with the following explanation: 

\begin{enumerate}
\item The \emph{first table}  is a fragment of the simple Cantor function. 
\item The \emph{second table}  illustrates the lexicographic ordering of finite sets of numbers in $\mathfrak{P}(\mathbb{N})$ according to $\phi_{||}(s) =   \pi((|s| ),\sigma_{|s|}(s))$ (equation \ref{CARD1}). Note the fact that we find the set $\{0,1,\dots,k\}$ on the line $y=0$ in column $k$. 
\item The \emph{third table} computes the shift for the sum function $\phi_{\Sigma}(s) = (\Sigma s,\theta_{\Sigma s}(s))$. 
\item The \emph{fourth table} computes the shift for multiplication $\phi_{\Pi}(s) = \pi(\Pi s,\theta_{\Pi s}(s))$.
\end{enumerate}

\subsection{A general theory of planar elastic translations for arithmetical functions}
This suggests a general construction for the study of the information efficiency of arithmetical functions:

\begin{definition}\label{SORTBIJ}
$\phi_{\zeta}:  \mathfrak{P}(\mathbb{N}) \rightarrow \mathbb{N}$, a \emph{injection sorted on $\zeta$}, is a mapping of the form:

\begin{equation}\label{GEN}
\phi_{\zeta}(s) = \pi(\zeta (s),\theta_{\zeta(s)}(s))
\end{equation}

where $\pi$ is the Cantor function and: 
\begin{itemize}
 \item $\zeta: \mathfrak{P}(\mathbb{N}) \rightarrow \mathbb{N}$ is a \emph{general arithmetical function operating on finite sets of numbers}. It can be interpreted as a type assignment function that assigns the elements of $\mathfrak{P}(\mathbb{N})$ to a type (column, sort) represented as a natural number.   
 \item   $\theta_k: \mathfrak{P}(\mathbb{N}) \rightarrow \mathbb{N}$ is an index function for each type $k$, that assigns an index to the set in column $k$.The equation $\theta_{\zeta(s)}(s)=n$ should be read as: $s$ is the $n$-th set for which $\zeta(s)=k$. 
\end{itemize}
\end{definition}

By theorem \ref{COUNTCARD} we have that $\phi_{||}$ is efficiently countable. We can use $\phi_{||}$  as a \emph{calibration device} to evaluate $\phi_{\zeta}$. If $\phi_{\zeta}$ is a sorted injection the following mappings exists $\phi_{||}^{-1}: \mathbb{N}  \rightarrow  \mathfrak{P}(\mathbb{N})$ and $\phi_{\zeta}^{-1}: \mathbb{N}  \rightarrow  \mathfrak{P}(\mathbb{N})$  such that $\phi_{||}^{-1}(\phi_{||}(x))$, $\phi_{\zeta}(\phi_{\zeta}^{-1}(x))$  are identities in $ \mathfrak{P}(\mathbb{N})$. Given this interconnectedness we can always use $\phi_{||}$ to construct $\phi_{\zeta}$:

\begin{algorithm}
\caption{Compute the sorted index function $\theta_{\zeta (s)}(s)$  using $\phi_{||}$.}
\begin{algorithmic}[1]
\STATE $ k \gets 0 $
\STATE $i \gets 0$
\WHILE {$k \leq \phi_{||}(s)$}
	\IF  {$\zeta (\phi_{||}^{-1}(k)) = \zeta(s)$}
	\STATE $k \gets k +1$
	\STATE $i \gets  i + 1 $ 
	\ELSE
	\STATE $k \gets k + 1 $
	\ENDIF
\ENDWHILE
\STATE $\theta_{\zeta (s)}(s) = i$  
\end{algorithmic}
\end{algorithm}

\begin{theorem}\label{PSEUDO}
If the function $\zeta$ exists and can be computed in polynomial time then sorted injections $\phi_{\zeta}:  \mathfrak{P}(\mathbb{N}) \rightarrow \mathbb{N}$ of the form $\phi_{\zeta}(s) = \pi(\zeta (s),\theta_{\zeta(s)}(s))$ exist and can be computed in time exponential to the representation of $s$.  
\end{theorem}
Proof: We have to compute $\phi_{\zeta}(s) = \pi(\zeta (s),\theta_{\zeta(s)}(s))$. The functions $\pi$ and $\zeta$ can be computed in polynomial time. The function $\theta$ can be computed using $\phi_{||}$ with algorithm 1.  This algorithm runs in time exponential in the representation of $k$ which is the index of the set $s$: $\phi_{||}^{-1}(k) = s$. 
$\Box$

Note that algorithm $1$ is a \emph{counting algorithm} as discussed in paragraph \ref{NonDetvsDet} and that  $\pi(\zeta (s),\theta_{\zeta(s)}(s))$ is a \emph{unique description} of $s$ as referred to in our central research question \ref{Q1}: Are there uniquely identifying descriptions of objects that contain more information than names of the objects they denote? Observe that this procedure is a meta-algorithm. It abstracts completely from the semantics of the function $\zeta$.

There is a spectrum of planar translations as is illustrated by table \ref{Spectrum}.  The interpretation of this table is as follows: 
\begin{enumerate}
\item The first column gives the definition of the \emph{elastic function classes with a planar representation} with increasing power: 
	\begin{enumerate}
		\item The \emph{Cantor function} with density $1$ in $\mathbb{N}^2$.
		\item \emph{Linear translations} by a factor $c$. 
		\item \emph{Polynomial translations} by $cx^k$. 
		\item \emph{Sum translations}, based on the sum of the set of natural numbers associated with the cell. 
		\item \emph{Product  translations}, based on the product of the set of natural numbers associated with the cell.
		\item The \emph{Trivial translation} defined by the function $\pi(x,y)=(\pi(x,y),0)$ that projects all the cells of the Cantor space on the line $y=0$.
 	\end{enumerate}, 

\item The second column gives the functions to compute the information efficiency of the translation over the $x$-axis.
\item The third column gives the number of resulting different efficiency functions in the limit. Constant for linear transformations. Growing unboundedly for polynomial translations, to a different efficiency function for each cell for the Sum and the Product translations. For the trivial translation $\pi$ defines itself as its own efficiency function. 
\item The fourth column  gives the number of different efficiency functions per cell. Only for the Sum and the Product translations this becomes larger than $1$. 
\item The fifth column gives the resulting density function for $\mathbb{N}^2$ over the counter diagonal. Up to polynomial translations the denstiy is $1$. For Sum and Product translations the bijection becomes an injection, because there are only a finite number of sets  that add up to, or multiply up to, a given number $n$. For the sum translation the density over the counter diagonal is roughly $x - \log x $ in the limit, for the  product translation the same density becomes logarithmic. For the trivial translation it becomes $0$. 
\end{enumerate}

\begin{table}[h!]
  \begin{center}
    \begin{tabular}{l|c|c|c|c} 
      \textbf{Translation} & \textbf{$\delta$ over} & \textbf{$\#$ Efficiency} &  \textbf{$\#$ per} &   \textbf{Density}\\
     \textbf{over $\mathbb{N}^2$} & \textbf{ $x$-axis} & \textbf{Functions } &  \textbf{cell} &    \textbf{in $\mathbb{N}^2$}\\
      \hline
      $\phi_{||}=\pi$ 	& $\delta(1)$ 	&$1$ 	& 1  & $1$ \\
      \hline   
            $\phi_{c} = \pi_{c}$	 &$ \delta(c)$ & $c$ & 1  & $1$ \\
      \hline   
      $\phi_{cx^k} = \pi_{cx^k}$	 &$ \delta(cx^k)$ & $\aleph_{0}$ & 1  & $1$ \\
      \hline
      $\phi_{\Sigma}$	&$\delta(\Sigma) $ &  $\aleph_{0}$ &  $> 2^x$  & $\approx x - \log x$ \\
      \hline
      $\phi_{\Pi}$		&$ \delta(\Pi)$ & $\aleph_{0}$  & $ \geq 1$ & $\approx \log x $ \\
      \hline
      $\phi_{\pi}$	&$ \delta(\pi)$ &$ 1 $  & 1 & $ \approx 0$ \\ \\
    \end{tabular}
    \caption{An overview of the spectrum of Cantor planar functions of increasing power. \label{Spectrum} }    
  \end{center}
\end{table}

 Metaphorically one would envisage the spectrum in terms of a rubber sheet lying over the space  $\mathbb{N}^2$. We measure the information efficiency resulting from the translations. When one starts to pull the sheet over the $x$-axis it starts to shrink over the $y$-axis. As long as one pulls with linear force the notion of a smooth information efficiency associated with a computable bijection over the space is conserved. When we pull with monotonely increasing force of a polynomial function, a computable bijection is still available, but the space starts to fluctuate with increasing discontinuities and the tension in the limit is infinite. When one pulls superpolynomially, using functions on sets of numbers, the rubber sheet starts to  disintegrate at finite dimensions and stops to be a computable bijection. It becomes an injection that is only occasionally locally computable on a point to point basis, although every stage of the translation can be reconstructed from the origin in exponential time.  When we pull the sheet super exponentially the shift is so big that the place of the image on the x-axis starts to carry information about the origin of all sets for which the image is on the line $x=c$. This is the case for primes and square numbers that have only two generating sets. When all points in the space $\mathbb{N}^2$ are mapped densely to the line $y=0$ the translation becomes the trivial inverse of the Cantor packing function, which is by definition information efficient again.

\subsection{A formal analysis of  the objects $\phi_{\Sigma}$ and $\phi_{\Pi}$}

Define $\Sigma s = \Sigma_{s_i \in s} s_i$ as the addition function for sets. Suppose  $s$ is the set associated with the index $(x,y)$ in $\mathbb{N}^2$ and $\pi(x,y)$ in $\mathbb{N}$. We have the sum function for sets: $\Sigma: \mathfrak{P}(\mathbb{N}) \rightarrow \mathbb{N}$.  The function $\epsilon_{\Sigma}: \mathbb{N}^2 \rightarrow \mathbb{N}^2$ defines an \emph{elastic translation} by a function $\Sigma$ of the form: 

\begin{equation}\label{SUMShift}
\epsilon_{\Sigma}(x,y) = (\Sigma s,\theta_{\Sigma s}(s))
\end{equation}

Here $\theta_{\Sigma}: \mathfrak{P}(\mathbb{N}) \rightarrow \mathbb{N}$ is the function that enumerates sets with the same sum. We have the functions: $\pi: \mathbb{N}^2 \rightarrow \mathbb{N}$ and $\phi_{||}: \mathfrak{P}(\mathbb{N}) \rightarrow \mathbb{N}$.  We define an injection $\phi_{\Sigma}:  \mathfrak{P}(\mathbb{N}) \rightarrow \mathbb{N}$ sorted on sum:

\begin{equation}\label{SUM}
\phi_{||}(\epsilon_{\Sigma}(s) )= \phi_{\Sigma}(s) = \pi(\Sigma s,\theta_{\Sigma s}(s))
\end{equation}

 A fragment of the function is shown in figure \ref{GRIDEXAMPLES} under the heading Sum Grid. The associated injection on $\mathbb{N}$ is: 
 \begin{equation}\label{B1X}
\begin{tikzcd}
\mathbb{N}\arrow{r}{\pi^{-1}}   & \mathbb{N}^2\arrow{r}{\epsilon_{\Sigma}} & \mathbb{N}^2\arrow{r}{\pi} & \mathbb{N}
\end{tikzcd} 
\end{equation}

 \begin{equation}\label{B2X}
 \begin{tikzcd}
\mathbb{N}\arrow{r}{\pi^{-1}}   & \mathbb{N}^2\arrow{r}{\epsilon_{\Sigma}^{-1}} & \mathbb{N}^2\arrow{r}{\pi} & \mathbb{N}
\end{tikzcd} 
\end{equation}
 
 Observe that the shift is an injection. For every natural number there is only a finite number of sets that add up to this number. The injection is dense when sampled over the counter diagonal.  

There are for each set $s_k = \{1,2,\dots  ,k\}$, with $\Sigma_{x \in s_k }  x = u$, according to equation \ref{Bell} exactly $B_k$ partitions that add up to $n$. This means that there is a super exponential number of sets that add up to $u$ which gives, when sampled over the counterdiagonal, using $\pi$, a density of $1$ in the limit. This fact is remarkable, at every line $x=c$ we compress an infinite set to a finite one, but the density of the resulting index stays close to $x - \log x$.
 
 \begin{figure}[!t]
\centering
\fbox{\includegraphics[ width=5in]{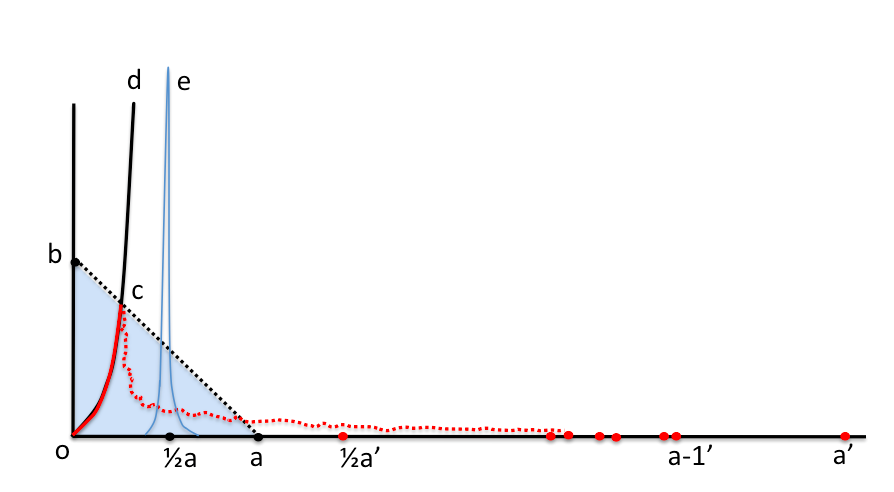}}
\caption{Schematic overview of the translation $\phi_{\Sigma}$ generated by $\phi_{||}$ at a finite moment $t$. \label{SUMtranslation}}
\end{figure}

In figure \ref{SUMtranslation} we give a schematic overview of the translation $\phi_{\Sigma}$ driven by the Cantor function for finite sets $\phi_{||}$ at a finite moment $t$. The area $(o,a,b)$ is the set of points that have been ``surveyed'' by $\phi_{||}$ up till now. The area below the red line illustrates the points that have been generated by  $\phi_{\Sigma}$. The exponential function illustrated by the line $(0,c,d)$ marks the empty area where $\phi_{\Sigma}$ will never generate images. Point $(a,0)$ is on line $y=0$. It codes the set $\{0,1,2,\dots a\}$ with the image $a'$ at $(\frac{1}{2}a(a-1),0)$. Point $a-1'$ is the image of $(a-1,0)$ at  $(\frac{1}{2}(a-1)(a-2),0)$. The gaussian distribution $e$ illustrates the density of the points around $\frac{1}{2}a$ that corresponds to sets of numbers, generated by a random selection of $\frac{1}{2}a$ elements from the set $\{0,1,2,\dots a\}$ that  will have an image in the neighborhood of $(\frac{1}{2} (\frac{1}{2}a (\frac{1}{2}a-1)),0)$. Note that only a tiny fragment of these sets has been surveyed by $\phi_{||}$. About al others we have no information at time $t$.

Observe that figure   \ref{SUMtranslation} describes the situation of the computation of  $\phi_{\Sigma}(s)$ at any finite moment in time and that the line $(a,b)$ is a hard \emph{information boundary}. At any moment in time $t$ the characterization of the set  $\phi_{\Sigma}$ by any finite segment of $\mathbb{N}$ is fundamentally incomplete. The images on the line $y=0$ form a discontinuous cloud of points. There is an exponential amount of relevant computations that we have not seen yet. We propose to call such sets \emph{Semi-Countable}. The name is motivated by the fact that these sets seem to occupy a place between countable and uncountable sets. There are no effective search procedures for such sets. These observations are formalized in the following theorem: 

\begin{theorem}\label{MAINSUM}
The function $\phi_{\Sigma}$ can: 1) be computed by a recursive function in time exponential in the representation of $s$ but 2) not in polynomial time. 
\end{theorem}
Proof:
\begin{enumerate}
\item $\phi_{\Sigma}(s)$ has a finite definition and for every set $s$ can be generated in time exponential to the representation of the Cantor index $\phi_{||}(s)$  according to theorem \ref{PSEUDO}. 

 \item  Suppose that there is an algorithm $p$ that computes  $\phi_{\Sigma}(s)$ in polynomial time. Observe that $p$ is either primitive recursive of $\mu$-recursive:  
\begin{itemize}

\item Algorithm $p$ cannot be a single  \emph{primitive recursive function} (constructive algorithm). This is an immediate consequence of the \emph{principle of characteristic information efficiency} (see definition \ref{InfEff}). 
There are two different grounds:

\begin{itemize}
\item The expression with a \emph{free variable} $x$ : 
\[\phi_{\Sigma}(x) = \pi(\Sigma x,\theta_{\Sigma x}(x))\] 
is a \emph{function scheme}. Each cell in $\mathbb{N}^2$ contains a different set $s$  Consequently $\phi_{\Sigma}(x)$ is a different primitive recursive computation for each $x=s$. Since $\Sigma s = k$ is information discarding the information about  the original computation is lost. In fact: since for every set $s$ the computation $\phi_{\Sigma}(x) = n$ is different, the number $\phi_{||}(s) = \pi(x)$ is an index of the computation. Consequently we can only compute $\phi_{\Sigma}^{-1}(n) = s$, if we already know $s$.

\item The expression with a \emph{bound variable}: \[\phi_{\Sigma}(s) = \pi(\Sigma s,\theta_{\Sigma s}s))\]
is also a \emph{function scheme}. Suppose $\Sigma s= u$ and $\theta_{\Sigma s}s=v$. The expression $(\Sigma s,\theta_{\Sigma s}s)$, is a \emph{unique description} of $s$: ``$s$ is  \emph{the $v$-th set that adds up to $u$''}. By lemma \ref{INFEFFADD} the information efficiency of this expression is not defined. This implies that the description is \emph{ad hoc}: it identifies $s$ uniquely, but it is not clear how much information we have when we see it.  By equation \ref{cNa} the number of possible different primitive recursive computations for $\Sigma$ is super-exponential and generates a cloud of different values for the information efficiency of each set. 
\end{itemize}
\item Algorithm $p$ cannot be a \emph{$\mu$-recursive function}  (search algorithm): The cell $(k,0)$ is projected on cell $(\frac{1}{2}k(k+1), 0)$, for all cells above $(k,0)$ the shift is larger. This means that the boundary conditions of theorem \ref{T1} apply.  $\phi_{\Sigma}$ generates an unbounded amount of information in the limit by equation \ref{INFGROW}. A polynomial time search algorithm can only generate  a constant amount of information. 
\end{itemize} 
\end{enumerate}
$\Box$

This implies that the denotation of the expression \emph{the $v$-th set that adds up to $u$''} is dependent on our search method. The search method itself creates the information about the objects it finds. Consequently there is no best or most effective way to search for \emph{the $v$-th set that adds up to $u$''}. This holds specifically for \emph{the first set that adds up to $u$''}.  The set $\phi_{\Sigma}$ cannot be searched effectively. 

The fact that the sum operation is information discarding illustrates the difference with the situation of theorem \ref{T1}. This is clarified by an analysis of the product translation. Observe that the product function is information efficient. We do not lose information when we multiply. This is an indication that $\phi_{\Pi}$ as a bijection is easier to compute than $\phi_{\Sigma}$.  Define $\Pi s = \Pi_{s_i \in s} s_i$ as the multiplication function for sets. We define an injection $\phi_{\Pi}:  \mathfrak{P}(\mathbb{N}) \rightarrow \mathbb{N}$ sorted on product: 

\begin{equation}\label{PROD}
\phi_{\Pi}(s) = \pi(\Pi s,\theta_{\Pi s}(s))
\end{equation}

\begin{figure}[!t]
\centering
\fbox{\includegraphics[ width=5in]{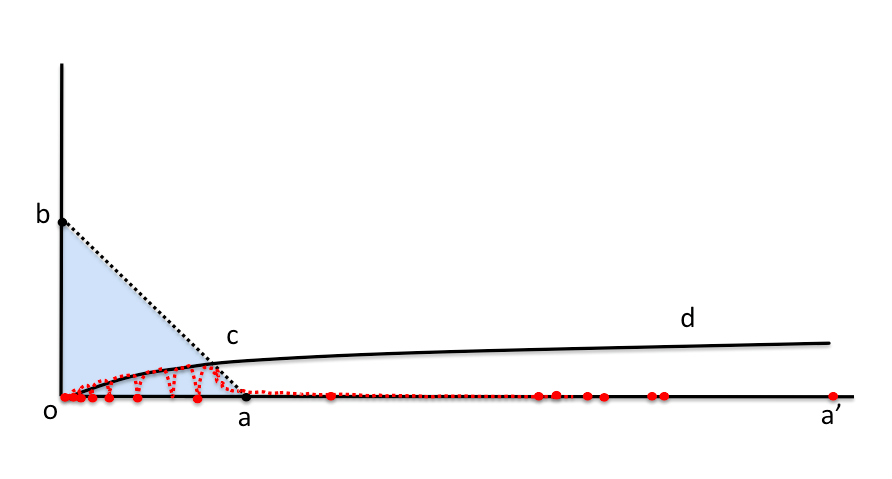}}
\caption{Schematic overview of the translation $\phi_{\Pi}$ generated by $\phi_{||}$ at a finite moment $t$. \label{PRODUCTtranslation}}
\end{figure}

A fragment of the function is shown in figure \ref{GRIDEXAMPLES} under the heading Product Grid. A schematic analysis of the translation is given in figure \ref{PRODUCTtranslation}. Observe that this shift is also an injection.The complexity of $\phi_{\Pi}$ is fundamentally different from the complexity of $\phi_{\Sigma}$.  We analyze the conditions of the proof of theorem \ref{MAINSUM} in the context of $\phi_{\Pi}(s)$:

 $\phi_{\Pi}(s)$ has a finite definition and for every set $s$ can be generated in time exponential to the representation of the Cantor index $\phi_{||}(s)$  according to theorem \ref{PSEUDO}. 
 
 Since multiplication is information efficient, we cannot invoke the \emph{principle of characteristic information efficiency} (see definition \ref{InfEff}) to prove that the is no recursive function to compute it. The information about  the original computation is kept in so far that the set of factors of $k$ always can be reconstructed. 

 For every natural number there is only a finite number of sets that multiply up to this number. There are for each set $s_k = \{1,2,\dots  ,k\}$, with $\Pi_{x \in s_k }  x = k!$, according to equation \ref{Bell} exactly $B_k$ partitions that multiply up to $k!$. This gives a density of the counter diagonal that is approximately logarithmic. 
 
 Since for every set $s$ the computation $\phi_{\Pi}(x) = n$ is different, the number $\phi_{||}(s) = \pi(x)$ is an index of the computation. Consequently we can only compute $\phi_{\Pi}^{-1}(n) = s$, if we already know $s$. This is exactly the case for prime numbers, they code the information about their own multiplicative history.  Since we know that primality is in $P$ the object  $\phi_{\Pi}$ is a bijection that at least partly for certain columns $x=c$ can be computed in polynomial time both ways. 
 
\subsection{The Subset Sum problem}

We give a detailed analysis of the conceptualization of the Subset Sum problem using $\phi_{\Sigma}$.  We get all possible instances of the subset sum problem when we generalize the sum operation over all possible subsets of $\mathbb{N}$. Let $f$ be the characteristic function of an infinite set $S \subseteq \mathbb{N}$. Example: the set $\{2,5,100,\dots \}$ is coded as $f(1)=2$, $f(2)=5$, $f(3)= 100$, $\dots$  Define $\Sigma_f (s) = \Sigma_{s_i \in s} f(s_i)$ as the addition function for sets of elements of $S$ ordered by their indexes in $\mathbb{N}$, e.g. $\Sigma_f (\{2,3\}) = f(2) + f(3) = 105$. We define an injection $\phi_{\Sigma, f}:  \mathfrak{P}(\mathbb{N}) \rightarrow \mathbb{N}$ sorted on sum of $S$:

\begin{equation}\label{SUM}
\phi_{\Sigma,f}(s) = \pi(\Sigma_f(s),\theta_{\Sigma_f (s)}s)
\end{equation}

Such injections generate gaps on the $x$-axis such that no point in the space $\mathbb{N}^2$ has an image on the line $x=c$. In these cases the existence of such an image becomes a \emph{decision problem}. Such decision problems can be computed in exponential time, but not in polynomial time: 

\begin{lemma}
$\phi_{\Sigma,f}(s)$ can: 1) be computed in time exponential in the representation of $s$ but 2) not in polynomial time. 
\end{lemma}
Proof: Suppose $\phi_{\Sigma,f}(s)$ can be solved in polynomial time. Observe that it  is a generalization of $\phi_{\Sigma}(s)$. We could solve $\phi_{\Sigma}(s)$ by choosing a specific $f$ and solve it. This contradicts theorem \ref{MAINSUM}.$\Box$

\section{Conclusion}
The objects $\phi_{\Sigma}$ and $\phi_{\Pi}$ belong to the most complex constructions we will ever encounter in mathematics. $\phi_{\Sigma}$ encodes everything there is to know about addition, while $\phi_{\Pi}$ encodes all we can ever know about multiplication including the locations of the primes. The class $\phi_{\Pi}$ is easier to compute than $\phi_{\Sigma}$. The last class is exactly in the sweet spot of elastic planar translation where the tension between compression (information loss on the $y$-axis)  and expansion (information generation over the $x$-axis) is maximal. Observe that any polynomial transformation of $\phi_{\Sigma}$ conserves the conditions of theorem \ref{MAINSUM}. Only when we apply exponential transformations, the deformations become so extreme that the translations start to carry information about the original cells in their new location on the $x$-axis and we come close to the easier class $\phi_{\Pi}$. 

We mention some remarkable observations that could motivate further study of the objects  $\phi_{\Sigma}$ and $\phi_{\Pi}$: 
\begin{itemize}
\item The Cantor function maps $\mathbb{N}$ onto $\mathbb{N}^2$. The objects  $\phi_{\Sigma}$ and $\phi_{\Pi}$ correspond to an infinite compression of $\mathbb{N}^2$ over the $y$-axis that still has a dense representation in $\mathbb{N}$ when ``sampled'' via the Cantor function. 
\item In terms of Kolmogorov complexity this implies that $\mathbb{N}$ is mapped onto itself without the \emph{expansion} that normally is generated by injective recursive functions like $y=x^2$ or $y=2^x$.  The images of the transalations in $\mathbb{N}$ generated by  $\phi_{\Sigma}$ and $\phi_{\Pi}$ are incompressible numbers. 
\item Consider figures \ref{Cantor_Shift}, \ref{SUMtranslation} and \ref{PRODUCTtranslation}. Note that all natural numbers  and measurements in these schematic geometrical images have logarithmic representations. Consequently any measurable neighborhood $\epsilon > 0$ has exponential size in the limit. We cannot search any area or distance with measure $>0$ efficiently in the limit. 
\item By the same reasoning any measurable neighborhood that is more than $\epsilon >0$ removed from $x=0$ or $y=0$ will contain an exponential number of  cells with incompressible indexes in the limit. 
 \item Observe figures \ref{SUMtranslation} and \ref{PRODUCTtranslation}. At any finite stage of the construction of these images there is an exponential  amount of points on the line $y=0$ for wich the corresponding image in the domain hes not been found. These cells correspond to the descriptions ``the first set $x$ that adds/multiplies up to  $k$. 
\end{itemize}

The semi-countable sets form an natural link between the countable and uncountable sets: $\mathbb{N}$ is countable, $\mathcal{P}(\mathbb{N})$ is uncountable. The class  $\mathfrak{P}(\mathbb{N})$ could be seen as the class that stretches  countability to the maximum. The elaborate scheme to map the class  $\mathfrak{P}(\mathbb{N})$ to $\mathbb{N}$ via the Cantor function and combinatorial number systems is in fact an application of the axiom of choice: we define an infinite set of different subclasses that allows us to make an infinite set of choices for the first element.  If we then define super polynomial information generating operations on such a mapping, the fabric of countability is torn apart and we enter the universe of semi-countable sets. 

\section{Acknowledgements}
I thank Peter van Emde Boas for many inspiring discussions and Rini Adriaans for continuous support. I thank the University of Amsterdam and especially Cees de Laat for allowing me the leeway to pursue my research interests.

\bibliographystyle{plain}
\bibliography{adriaans}

\begin{thebibliography}{1}

\bibitem{AB2011}
P.W. Adriaans and P.E. {Van Emde Boas}.
\newblock Information, computation and the arrow of time.
\newblock In B.~S. Cooper and A.~Sorbi, editors, {\em Computation and Logic in
  the Real World}, COMPUTABILITY IN CONTEXT, chapter~I. Imperial College Press,
  2011.

\bibitem{FP23}
R.~Fueter and G.~P\'{o}lya.
\newblock Rationale {A}bz\"{a}hlung der {G}itterpunkte.
\newblock {\em Vierteljschr. Naturforsch. Ges. Z\"{u}rich}, 58:280--386, 1923.

\bibitem{LiVi08}
M.~Li and P.~Vit\'{a}nyi.
\newblock {\em An Introduction to Kolmogorov Complexity and Its Applications}.
\newblock Springer-Verlag, 2008.

\bibitem{Odi16}
P.~Odifreddi and B.S. Cooper.
\newblock Recursive functions.
\newblock In E.N. Zalta, editor, {\em The Stanford Encyclopedia of Philosophy}.
  Metaphysics Research Lab, Stanford University, fall 2016 edition, 2016.

\bibitem{adri2018-1}
P.W.Adriaans.
\newblock Information.
\newblock In E.N. Zalta, editor, {\em The Stanford Encyclopedia of Philosophy.}
  Metaphysics Research Lab, Stanford University, fall 2018 edition, 2018.

\end{thebibliography}

\newpage

\section{Appendix: Combinatorial Considerations\label{A3}}
The product of a set of numbers $s$ is:  $\Pi_{i \in s} S_i$. The sum of a set of numbers $s$ is: $\Sigma_{i \in s} s_i$. The set of numbers $\{1,2,\dots n\}$ is called an \emph{initial segment} of $\mathbb{N}$ of size $i$.  
The sum of an initial segment $\{1,2,\dots n\}$ is:  
\begin{equation}\label{SumSeg}
\Sigma_{i=1}^n i = \frac{1}{2}n(n+1)
\end{equation}
The product of an initial segment $\{1,2,\dots n\}$ is:  
\begin{equation}\label{ProductSeg}
\Pi_{i=1}^n i = n!
\end{equation}
The number of ways to write down $n$ pairs of balanced brackets is given by the \emph{Catalan} number: 
\begin{equation}\label{Catalan}
C_n = \frac{1}{n-1}{2n \choose n }
\end{equation}
The \emph{Stirling number of the second kind} is the number of ways a set of $n$ numbers can be partioned into $k$ non-empty subsets: 
\begin{equation}\label{Stirling}
\stirling{a}{b}= \frac{1}{k!} \Sigma_{i=0}^{k} (-1)î {k \choose i }(k-i)^n
\end{equation}
The \emph{Bell} number counts the number of possible partitions of a set: 
\begin{equation}\label{Bell}
B_n= \Sigma_{k=0}^{n}\stirling{n}{k} 
\end{equation}

Let $\oplus_{\not a, \not c}: \mathbb{N}^2 \rightarrow \mathbb{N}$ be a tensor operator that is \emph{non-commutative} and {non-associative}.  Suppose $\# \oplus_{\not a, \not c}$ is the number of different computations we can perform with $\oplus$ on a set of $n$ numbers. It is given by the number of ways we can write $n-1$ pairs of brackets in $n!$ different sequences of strings:

\begin{equation}\label{NcNa}
\#\oplus_{\not a, \not c}(n) = n!C_{n-1} = n! \frac{1}{n}{2(n-1) \choose (n-1) }
\end{equation}
Suppose $\oplus_{\not a, c}$ is \emph{commutative} and \emph{non-associative}, then, to our knowlegde, there is no formula that describes the number of different computations for $n$ objects, but it is, for large enough $n$, certainly bigger than the total number of unbracketed partitions, which is given by the Bell number, and thus  smaller than $\#\oplus_{\not a, \not c}(n)$ but still super exponential: 
\begin{equation}\label{cNa}
\#\oplus_{\not a, \not c}(n) > \#\oplus_{\not a, c}(n) > B_n > 2^n
\end{equation}

If $\oplus_{a, c}$ is \emph{commutative} and \emph{associative}, then the number of different computations for $n$ objects is $1$: all computations can be transformed in to all others. Which gives the following correlation between structural rules and number of computations defined on sets: 
\begin{equation}\label{cNa}
\#\oplus_{\not a, \not c}(n) = n!C_{n-1}  > \#\oplus_{\not a, c}(n) > B_n > 2^n > \#\oplus_{a, c}(n) = 1
\end{equation}

\section{Appendix:	 Information Efficiency of  Recursive Functions \label{A1}} 

We have defined $\forall (x \in \mathbb{N}) I(x)=\log x$. Note that $n \in \mathbb{N}$ is a \emph{cardinality} and that the measurements $I(n)$ represent a cloud of \emph{scalar} point values in $\mathbb{R}$. The relation between the two sets is described by the Taylor series for $\log (x+1)$:   
\begin{equation}\label{MEASINFGROW}
I(s+1) = \log (x + 1) =  \Sigma_{n=1}^{\infty}(-1)^{n-1}\frac{x^n}{n} \end{equation}

We have $\lim_{x \rightarrow \infty} I(x+1) - I(x) = 0$. In the limit the production of information by counting processes stops as specified by the derivative $\frac{d}{dx} \ln x = \frac{1}{x}$. The intuition that one could express based on this observation is that information in numbers in the limit becomes extremely dense, much denser than the values in `normal' counting processes. 

Given definition \ref{EFFFUNCTION}. We can construct a theory about the flow of information in computation. For  primitive recursive functions we follow \cite{Odi16}.  Given  $\log 0 = 0 $ the numbers $0$ and $1$ contain no information. 

\begin{itemize}
\item \emph{Composition of functions} is information neutral: 
\[\delta(f(g(x))) + \delta(g(x)) = \log f(g(x)) - \log g(x) + \log g(x) - \log x =\]\[ \log f(g(x)) - \log x\]

\item  The \emph{constant function} $z(n)=0$ carries no information $z^{-1}z(n)= \mathbb{N}$.  
\item The first application of the \emph{successor function} $s(n)=x+1$ is information conserving:  \[\delta(s(0))=  \log (0 + 1) - \log 0 = 0\] 
\item The \emph{successor function} expands information for values $>1$. By equation \ref{MEASINFGROW} we have:   
\[ I(s) = \log (x + 1) =  \Sigma_{n=1}^{\infty}(-1)^{n-1}\frac{x^n}{n} > \log x \]
Consequently: 
  \[\delta(s(x))= I(s(x)) - \log x - \log 1=  \log (x + 1) -\log x = \epsilon > 0\]
  Note that $\epsilon$ goes to zero as $x$ goes to $\infty$. 
\item The \emph{projection function} $P_{i,n}((x_1,x_2,\dots,x_n) = x_i$, which returns the i-th argument $x_i$,  is information discarding. Note that the combination of the index $i$ and the ordered set $(x_1,x_2,.., x_n)$ already specifies $x_i$ so: 
 \[\delta(P_{i,n}(x_1,x_2,.., x_n)=  I(x_i) - \log i - I(x_1,x_2,\dots, x_n) <  0\]  
\item \emph{Substitution.} If $g$ is a function of $m$ arguments, and each of $h_1,\dots,h_m$  is a function of $n$ arguments, then the function $f$:
\[f(x_1,\dots,x_n)=g(h_1(x_1,\dots,x_n),\dots,h_m(x_1,\dots,x_n))\]
is definable by composition from $g$ and $h_1,\dots,h_m$. We write $f=[g \circ h_1,\dots,h_m]$, and in the simple case where $m=1$  and $h_1$  is designated $h$, we write $f(x)=[g \circ h](x)$. Substitution is information neutral: 
\[\delta( f(x_1,\dots,x_n))=\delta(g(h_1(x_1,\dots,x_n),\dots,h_m(x_1,\dots,x_n))) =\]
\[I(f(x_1,\dots,x_n) - I(x_1,\dots,x_n) \]
Where $I(f(x_1,\dots,x_n)$ is dependent on $\delta(g)$ and $\delta(h)$. 
\item \emph{Primitive Recursion.} A function $f$ is definable by primitive recursion from $g$ and $h$ if $f(x,0)= g(x)$ and $f(x,s(y))=h(x,y,f(x,y))$.  Primitive recursion is information neutral: 
\[\delta(f(x,0))= \delta(g(x)) = I(g(x) - I(x)\]
which dependent on $ I(g(x)$ and 
\[\delta(f(x,s(y)))=\delta(h(x,y,f(x,y))) = I(h(x,y,f(x,y))) - I(x) - I(y)\]
which is dependent on $I(h(x,y,f(x,y)))$.
 \end{itemize}  
  
 Summarizing: the primitive recursive functions have one information expanding operation, \emph{counting}, one information discarding operation, \emph{choosing}, all the others are information neutral.

 The information efficiency of more complex operations is defined by a combination of counting and choosing. We analyse  the definition of addition by primitive recursion:  suppose $f(x) = P^1_1(x) = x$ and $g(x,y,z)= S(P_2^3(x,y,z)) = S(y)$. Then $h(0,x) = x$ and $h(S(y),x) = g(y,h(y,x),x) = S(h(y,x))$. The information efficiency is dependent on the balance between choosing and counting in the definition of $g$: 
  \[\delta(g(x,y,z))=  I(S(y)) - I(x,y,z) = (\Sigma_{n=1}^{\infty}(-1)^{n-1}\frac{y^n}{n}) - I(x,y,z) \]
  Note that the operation looses information and the information loss is asymmetrical. The commutativity and associativity of information efficiency for complex arithmetical operations are not trivial:   
  
\begin{theorem}[Information Efficiency of Elementary Arithmetical Operations]\label{BASICRULES}
\begin{itemize}  
\item \emph{Addition of different variables is information discarding}.  In the case of addition we know the total number of times the successor operation has been applied to both elements of the domain: for the number $c$ our input is restricted to the tuples of numbers that satisfy the equation $a + b =c $ $(a,b,c \in \mathbb{N})$. Addition is information discarding for numbers $>2$: 
 \begin{equation}\label{EA1}
 \delta(x + y)= \log (x + y) -\log x - \log y < 0 
 \end{equation}
 
\item \emph{Addition of the same variable has constant information}. It measures the reduction of information in the input of the function as a constant term:
\begin{equation}\label{EA2}
\forall (x) \delta(x + x)=  \log 2x -\log x  = \log 2 
\end{equation}

\item \emph{Commutativity for addition is information efficient:} 
\begin{equation}\label{COMMADD}
\delta (x + y) = \log (x + y) -\log x - \log y = \log (y + x) -\log y - \log x  = \delta (y + x)
\end{equation}

\item \emph{Associativity for addition is not information efficient:} 
\begin{equation}\label{ASSADD}
\begin{split}
\delta(x + (y + z)) = \log (x + (y + z) ) -\log x - \log (y + z) \neq \\\log ((x + y) + z ) -\log (x + y) - \log z =  \delta ((x + y) + z)
\end{split}
\end{equation}
  \item  \emph{Multiplication of different variables is information conserving.} In the case of multiplication  $a \times b =c $ $(a,b,c \in \mathbb{N})$ the set of tuples that satisfy this equation is much smaller than for addition and thus we can say that multiplication carries more information. If $a=b$ is prime (excluding the number $1$) then the equation even identifies the tuple.  Multiplication is information conserving for numbers $>2$: 
\begin{equation}\label{EA3}
 \delta(x \times y) = \log (x \times y) -\log x - \log y = 0 \end{equation} 
 
 \item \emph{Multiplication by the same variable is information expanding}. It measures the reduction of information in the input of the function as a logarithmic term:
\begin{equation}\label{EA4}
\forall (x) \delta(x \times x) = \log (x \times x) -\log x  = \log x > 0
\end{equation}

\item \emph{Commutative for multiplication is information efficient:} 
\begin{equation}\label{COMMUL}
\delta(x \times y) = \log (x \times y) -\log x - \log y =   \log (y \times x) -\log y - \log x = \delta(y \times x)
\end{equation}

\item \emph{Associativity for multiplication is information efficient:}  
\begin{equation}\label{ASSMUL}
\begin{split}
\delta(x \times (y \times z)) = \log (x \times (y \times z) ) -\log x - \log (y \times z) = \\ \log ((x \times y) + z ) -\log (x \times y) - \log z =  \delta ((x \times y) \times z)
\end{split}
\end{equation}
\end{itemize}

\end{theorem}

Estimating the information efficiency of elementary functions is not trivial. From an information efficiency point of view the elementary arithmetical functions are complex families of functions that describe computations with the same outcome, but with different computational histories. For addition the effect is significant:

\begin{lemma}\label{INFEFFADD}
If $s$ is a set of natural numbers then $\delta(\Sigma_{1 \in s}s_i)$ is not defined.
\end{lemma}
Proof: Immediate consequence of theorem \ref{BASICRULES}: the information efficiency of addition is non-associative. By equation \ref{cNa} the number of possible different computations is super-exponential in the cardinality of the set and generates a cloud of different values for the information efficiency of each set. $\Box$

As an illustration we work out the example given in the introduction (see par. \ref{MA}). Suppose we want to add the numbers $2$, $47$, $53$ and $98$. Most of us will see the pattern that makes this easy: $(2+98) + (47 + 53) = 100 + 100 = 200$. The numbers have a special relationship that makes the result less surprising, and therefore less informative. Apart from that the conditional information in the number $200$  given the set $\{2, 47, 53, 98\}$ varies fundamentally with the method of computation:   

\begin{itemize}
\item As a composed function of $4$ variables:

\begin{equation} \label{EXEFF1}
2 + 47 + 53 + 98 = 200
\end{equation}

\[\delta(2 + 47 + 53 + 98) = \]
\[\log 200 - \log 2 - \log 47 - \log 53 - \log 98 \approx  \]
\[7,64 - 1 – 5,56 - 6,64 - 6,61 =\]
Total:  $-12,17$

\item Storing intermediate results, method $1$: 

\begin{equation}\label{EXEFF2}
(2 + 47) + (53 + 98) = 200
\end{equation}

\[\delta(2 + 47) = \log 49 - \log 2 - \log 47 \approx 
5,61 - 1 – 5,56 = -0.95\]
\[\delta(53 + 98) = \log 151 - \log 53 - \log 98 \approx 
7,24 – 5,56 – 5,78  = - 4,1\]
\[\delta(49 + 151) = \log 200  - \log 49 - \log 151 \approx 
7,64 – 5,61 – 7,24 = - 5,21\]
\[- 0,95 - 4,1 - 5,21 =\]
Total: $-10,26$

\item Storing intermediate results, method $2$: 

\begin{equation}\label{EXEFF3}
(2 + 98 )+ (47 + 53) = 200
\end{equation}
\[\delta(2 + 98) = \log 100 - \log 2 - \log 98 \approx  
6,64 - 1 - 6,61 = -0.97 \]
\[\delta(47+ 53) =  \log 100 - \log 47 - \log 53 \approx 
6,64 - 5,56 - 5,78 = – 4,70 \]
\[\delta(100 + 100) = \log 200 - \log 100 - \log 100 \approx
7,64 - 6,64 - 6,64 = -5,64\]
\[- 0,97 - 4,70  - 5,64=\]
Total: $ - 11,31$

\item Storing intermediate results, method $2$, and applying equation \ref{EA2}:

\begin{equation}\label{EXEFF4}
\delta(100 + 100) = \log_2 2 = 1
\end{equation}
Total: $- 0.97 - 4.70 + 1 = -4,67$
\end{itemize}

Observe that basic composed equation \ref{EXEFF1} is not the most information efficient one with value $-12,17$. Storing intermediate results actually improves the information efficiency in equations \ref{EXEFF2} with value  $-10,26$ and \ref{EXEFF4} with value $- 11,31$.  The latter is somewhat less efficient, a situation that changes when we apply equation \ref{EA2}. 

The associativity of information efficiency for multiplication is an `accident' caused by the information conservation of the operation, but the effect of collapse of arguments as in equation \ref{EA4} remains.  Summarizing: some arithmetical operations expand information, some have constant information and some discard information. The flow of information is determined by the succession of types of operations, and by the balance between the complexity of the operations and the number of variables.

\section{Appendix: Computing the bijection between $\mathfrak{P}(\mathbb{N})$ and $\mathbb{N}$ via $\mathbb{N}^2$ using the Cantor packing function: an example \label{A2}} 

\subsection{From $\mathfrak{P}(\mathbb{N})$ to $\mathbb{N}$}
Take a random finite set of natural numbers:  
\[\{0,3,5,7,9,10\}\]

Convert to $\mathbb{N}^+$: 
\[1,4,6,8,10,11\]

\subsubsection{Compute the index in a combinatorial number system of order $6$}

\[\sigma_6(s)=  {11\choose 6} + {10 \choose 5}+ {8 \choose 4} + {6 \choose 3}+ {4 \choose 2} + {1 \choose 1} =\]

\[462 + 252 + 70 + 20 + 6 + 1 = 811\]

\subsubsection{Compute the Cantor index for  location $(6,811)$}

\[\pi(6,811) = 1/2 ((6 + 811  + 1) (6 + 811 )) + 811 = 334964\]

 \subsection{From $\mathbb{N}$ to $\mathfrak{P}(\mathbb{N})$}

Take the  index: $\pi(x,y) = 334964= z$. 

 \subsubsection{Compute the location for $334964$}

\[\pi(x,y)=1/2(x+y+1)(x+y)+y\]

Define\footnote{We follow the computation given in https://en.wikipedia.org/wiki/Pairing\_function, retrieved February 27, 2018. }.

\[w = x + y\]

\[t = \frac{w(w+1)}{2} = \frac{w^2 + w}{2}\]

\[z = t+y\]

Here $t$  is the triangular number of  $w$. We solve: 

\[w^2 + w -2t = 0\]

\[w = \frac{\sqrt{8 t + 1} - 1 }{2}\]

which is a strictly increasing and continuous function when $t$ is a non-negative real. Since

\[t \leq z =t+y < t+(w+1) = \frac{(w+1)^2 + (w+1)}{2}\]

we get that 

\[w \leq  \frac{\sqrt{ 8 z + 1} - 1}{2} < w + 1\]

and thus

\[w = \lfloor \frac{\sqrt{ (8 \times 334964) + 1} - 1}{2}\rfloor = 817\]

\[t = (817^2 + 817)/2 = 334153\]

\[y = z – t\]

\[y = 334964 - 334153 = 811\]

\[x = 817 - 811 = 6  \]

The set has cardinality  $6$ and index $811$. 

\subsubsection{Compute the set associated with location $(6,811)$}

We reconstruct the number sequence from $\mathbb{N}^+$. Compute: ${x \choose 6} < 811 = 11.802 \dots$. The biggest number is $11$ 

\[811 -  {11 \choose 6} = 349 \]

Compute: ${x \choose 5} < 349 = 10.521 \dots$

Next number  is $10$. 

\[349 -  {10 \choose 5} = 97\]

\[97 - {8 \choose 4} = 27\]

\[27 - {6 \choose 3} = 7\]

\[7 - {4 \choose 2} = 1\]

\[{1 \choose 1} = 1\]

The sequence is $1,4,6,8,10,11$. Convert to $\mathbb{N}$. The original set  is $\{0,3,5,7,9,10\}$

\end{document}